\documentclass[12pt,aps,prd,
nofootinbib,floatfix,superscriptaddress,a4paper]{revtex4}

\usepackage{graphics}
\usepackage{epsfig}
\usepackage{amsmath,amsfonts,amssymb}
\usepackage{color}
\usepackage{epsf}
\usepackage{float}
\usepackage{subfigure}
\newcommand {\ignore}[1]{}
\newcommand{\eVq}  {\text{eV}^2}

\def\lsim{\raise0.3ex\hbox{$\;<$\kern-0.75em\raise-1.1ex\hbox{$\sim\;$}}}
\def\gsim{\raise0.3ex\hbox{$\;>$\kern-0.75em\raise-1.1ex\hbox{$\sim\;$}}}

\newcommand{\AddrSIBA}{%
Department of Physics, Visva-Bharati,
Santiniketan - 731 235, India}

\newcommand{\AddrAHEP}{%
  AHEP Group, Institut de F\'{\i}sica Corpuscular --
  C.S.I.C./Universitat de Val{\`e}ncia \\
  Edificio Institutos de Paterna, Apt 22085, E--46071 Valencia, Spain}
  
\newcommand{\AddrGranada}{%
  Departamento de F\'isica Te\'orica y del Cosmos and CAFPE,\\
Universidad de Granada, E-18071 Granada, Spain}

\newcommand{\AddrUCL}{%
 Department of Physics and Astronomy, University College London,\\
London WC1E 6BT, United Kingdom}

\def\321{SU(3) $\otimes$ SU(2) $\otimes$ U(1)}
\def\lr{SU(2)$_L~\otimes$ SU(2)$_R~\otimes$ U(1)$_{B-L}$ }

\begin{document}


\title{
   Heavy Neutrinos and Lepton Flavour Violation\\
   in Left-Right Symmetric Models at the LHC
}                       

\author{S.~P.~Das}      \email{spdas@ific.uv.es}             \affiliation{\AddrSIBA}
\author{F.~F.~Deppisch} \email{f.deppisch@ucl.ac.uk}         \affiliation{\AddrUCL}
\author{O.~Kittel}      \email{kittel@th.physik.uni-bonn.de} \affiliation{\AddrGranada}
\author{J.~W.~F.~Valle} \email{valle@ific.uv.es}             \affiliation{\AddrAHEP}

\begin{abstract}
\noindent
We discuss lepton flavour violating processes induced in the production and decay of heavy right-handed neutrinos at the LHC. Such particles appear in left-right symmetrical extensions of the Standard Model as the messengers of neutrino mass generation, and can have masses at the TeV scale. We determine the expected sensitivity on the right-handed neutrino mixing matrix, as well as on the right-handed gauge boson and heavy neutrino masses. By comparing the sensitivity of the LHC with that of searches for low energy LFV processes, we identify favourable areas of the parameter space to explore the complementarity between LFV at low and high energies.
\end{abstract}


\maketitle

\newpage


\section{Introduction}
\label{sec:Introduction}

The discovery of neutrino oscillations~\cite{fukuda:1998mi, ahmad:2002jz, eguchi:2002dm} shows that neutrinos are massive~\cite{Maltoni:2004ei} and that lepton flavour is violated in neutrino propagation. It is natural to expect that the violation of this conservation law should show up in other contexts, such as rare lepton flavour violating (LFV) decays of muons and taus, e.g. $\mu^-\to e^-\gamma$, and possibly also at the high energies accessible at the Large Hadron Collider (LHC).
In addition to proving that flavour is violated in the leptonic sector, oscillation experiments have convincingly shown that at least two of the three active neutrinos have a finite mass. However, despite this success, oscillation experiments are unable to determine the absolute magnitude of neutrino masses. 
Altogether there are three complementary approaches to probe the absolute scale of neutrino mass.
Upper limits on the effective electron neutrino mass of $\sim$~2~eV can be set from the analysis of tritium beta-decay experiments~\cite{kraus:2004zw, osipowicz:2001sq}. Astronomical observations combined with cosmological considerations also allow an upper bound to be set on the sum of the three neutrino masses of the order of 0.7~eV, under some assumptions~\cite{Komatsu:2008hk}.
The observation of neutrinoless double beta decay ($0\nu\beta\beta$) would signal the Majorana nature of neutrinos and the violation of total lepton number~\cite{schechter:1982bd, Duerr:2011zd}.
Corresponding searches provide an upper bound on the $0\nu\beta\beta$ effective Majorana neutrino mass parameter $m_{\beta\beta} \leq 300-600$~meV~\cite{KlapdorKleingrothaus:2000sn}.
Again, it is to be expected that, if $0\nu\beta\beta$ is observed, the violation of total lepton number should also, at some level, take place at high energy accelerators like the LHC.

Several mechanisms of neutrino mass generation have been suggested in the literature, the most prominent example being the seesaw mechanism in which heavy right-handed Majorana neutrinos act as ``messengers'' by generating light Majorana masses for the observed active neutrinos through their mixing with the left-handed neutrinos. The Majorana character of the active neutrinos can then be connected to a breaking of lepton number symmetry at a scale possibly associated with unification~\cite{minkowski:1977sc, gell-mann:1980vs, yanagida:1979, mohapatra:1979ia, schechter:1980gr, schechter:1981cv, lazarides:1980nt} and might also be responsible for the baryon asymmetry of the Universe through the leptogenesis mechanism~\cite{fukugita:1986hr}.

Despite its attractiveness, the default type-I seesaw mechanism has important phenomenological shortcomings: In the standard regime, the right-handed neutrinos have masses close to the unification scale and can therefore not be directly produced. 
In addition, the right-handed neutrinos are gauge singlets. This means that even if the masses are low enough for them to be produced, the heavy neutrinos only couple with Yukawa strength, tightly constrained by the smallness of neutrino masses\footnote{The mixing of the heavy neutrino singlets is also constrained by precision data such as the rates for lepton flavour violating processes \cite{Forero:2011pc, Abada:2007ux, delAguila:2008pw}.}. This implies that the simplest seesaw schemes are difficult to test at the LHC~\cite{delAguila:2007em}.

A widely studied alternative of the standard Seesaw scheme with gauge singlet heavy neutrinos is the left-right symmetrical model (LRSM) which extends the electroweak Standard Model gauge symmetry SU(2)$_L\,\,\otimes$ U(1) to the \lr group~\cite{Pati:1974yy, Mohapatra:1974gc, Senjanovic:1975rk, Duka:1999uc}. Here, right-handed neutrinos are necessary to realize the extended gauge symmetry and come as part of an SU(2)$_R$ doublet, coupling to the heavy gauge bosons. As a result, heavy neutrinos can be produced with gauge coupling strength, with promising discovery prospects, given the relatively weak direct experimental bounds on the masses of the extra gauge bosons.

\begin{figure}[t]
\subfigure[]
{\label{fig:diagramsLHCa}\includegraphics[clip,width=0.47\textwidth]{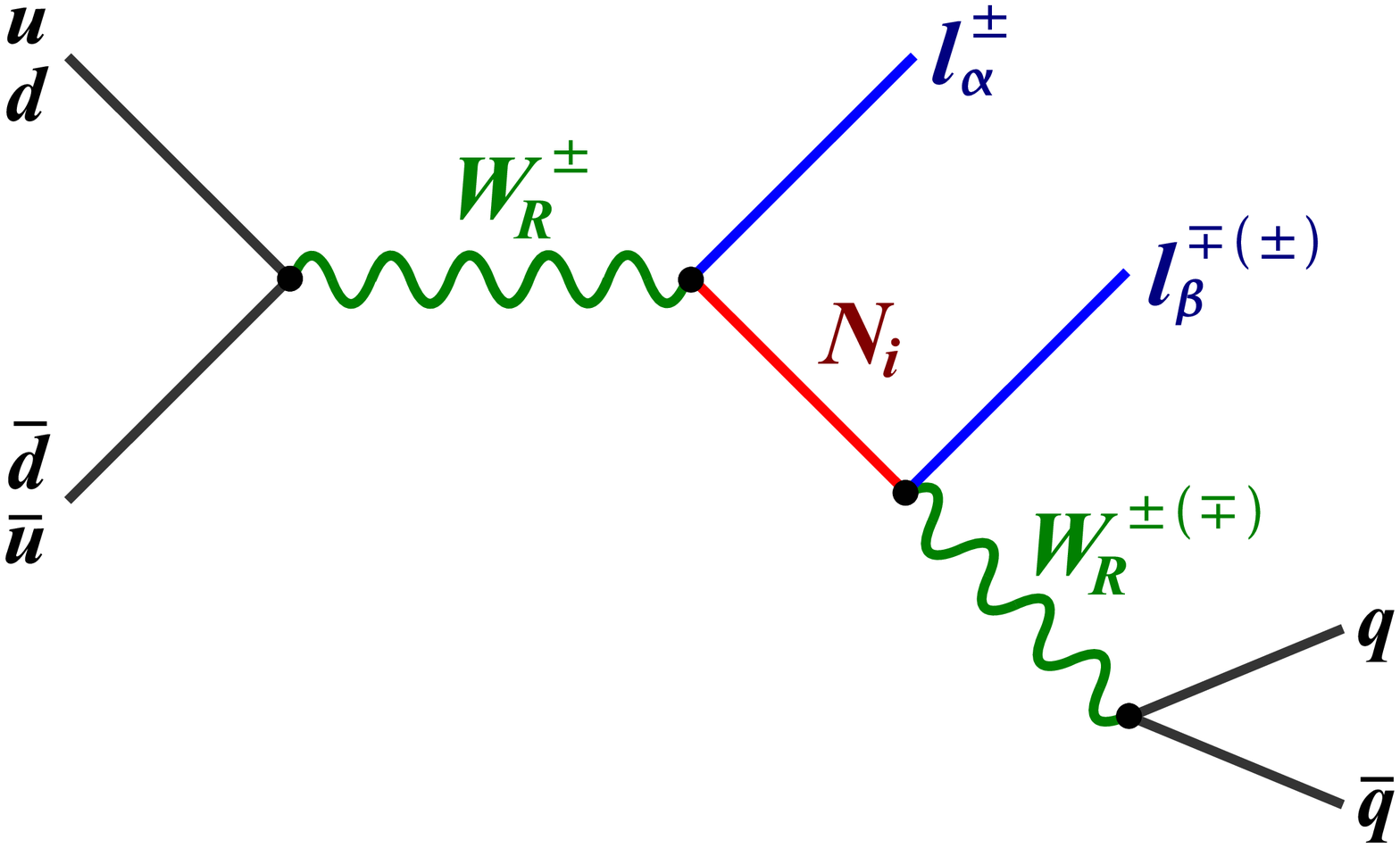}}
\subfigure[]
{\label{fig:diagramsLHCb}\includegraphics[clip,width=0.47\textwidth]{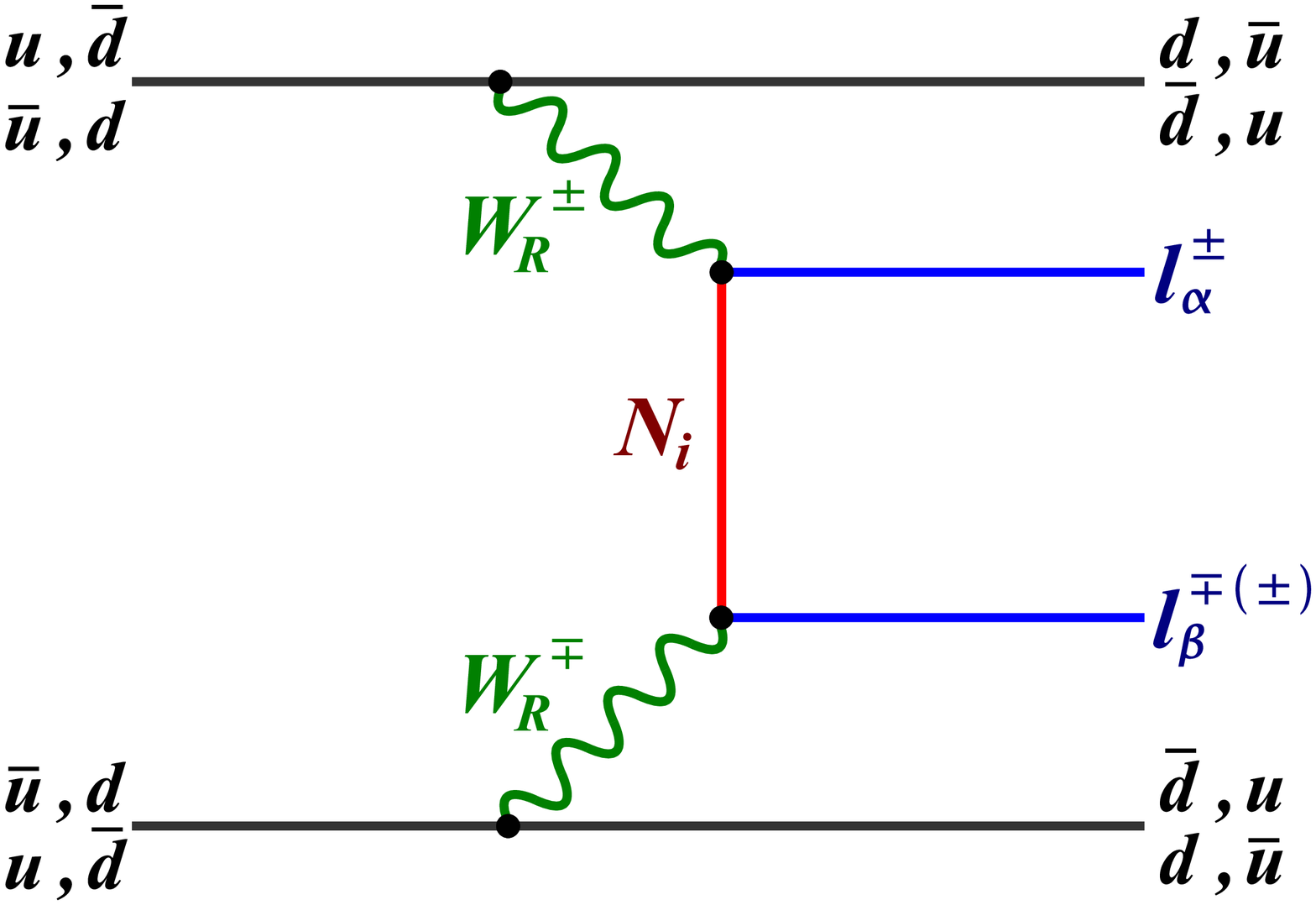}}
\\
\subfigure[]
{\label{fig:diagramsLHCc}\includegraphics[clip,width=0.47\textwidth]{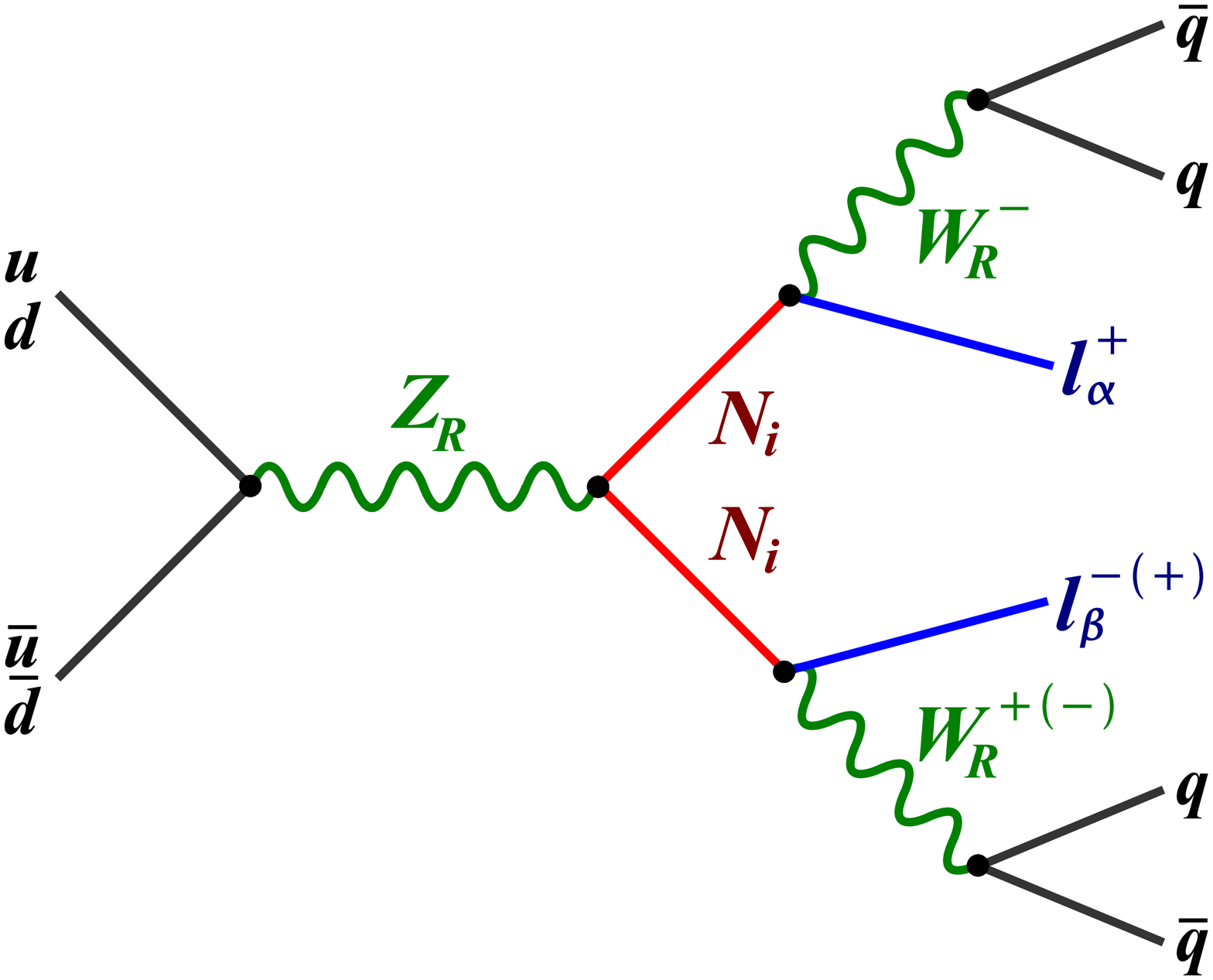}}
\caption{Different production processes of heavy right-handed neutrinos with dilepton signatures at hadron colliders.}
\label{fig:diagramsLHC}
\end{figure}

In this paper we address the prospects of probing the mass matrix of the right-handed neutrinos. With the existing information on the light neutrino sector and low energy lepton flavour violation, we re-derive the expected LHC sensitivities on the right-handed gauge boson and heavy neutrino masses. Among the possible production processes shown in Figure~\ref{fig:diagramsLHC}, we will focus on the resonant production of a right-handed $W_R$ boson displayed in Figure~\ref{fig:diagramsLHCa}. In addition, we determine the sensitivity of lepton flavour violating processes at the LHC  to measure the right-handed neutrino masses and their mixing, and compare results with low energy LFV searches. The focus in this paper is on lepton flavour violating effects rather than lepton number violation, and we will take into account both opposite-sign and same-sign lepton events of the process in Figure~\ref{fig:diagramsLHCa} at the LHC. Nevertheless, we will briefly comment on dedicated searches of lepton number violation using same-sign event signatures and the interplay with $0\nu\beta\beta$.

This paper is organized as follows. In Section~\ref{sec:model} we introduce the minimal LR model, and discuss the structure of its charged current couplings. Section~\ref{sec:limits} covers the main observables which give current limits on the LR model used. The dilepton signals arising from heavy neutrino production are discussed in Section~\ref{sec:ResultsDilepton}. Finally, we present our conclusions in Section~\ref{sec:conclusion}.

\section{Left-Right Symmetry}
\label{sec:model}

A striking feature of neutrino oscillation data is that the leptonic mixing angles follow a pattern substantially different from that which characterizes quarks~\cite{nunokawa:2007qh}. Within the seesaw mechanism, as implemented in left-right symmetrical models, neutrino masses are generated in a different way from charged fermion masses, as a result of the violation of lepton number. Hence, in principle, one may reconcile the large solar and atmospheric angles indicated by neutrino oscillation data with the small Kobayashi-Maskawa angles. Many specific models based on flavour symmetries have been suggested for this~\cite{Hirsch:2012ym}. In what follows we will simply assume a generic lepton flavour structure with the right-handed neutrinos and the charged gauge bosons of the left-right model in the range of a few TeV.

\subsection{Minimal Left-Right Symmetrical Model}
\label{sec:minimalLR}

In the minimal \lr model with manifest left-right symmetry, a generation of quarks and leptons is assigned to the multiplets~\cite{Barenboim:1996vu}
\begin{equation}
	Q_{L,R}=
	\begin{pmatrix}
		u \cr d \cr
	\end{pmatrix}_{L,R}, \quad
	\psi_{L,R}=
	\begin{pmatrix}
		\nu \cr \ell \cr
	\end{pmatrix}_{L,R},
\end{equation}
with the quantum numbers under \lr
\begin{eqnarray}
	Q_L: \left(\frac{1}{2},0,\frac{1}{3} \right), 
	\quad \psi_L: \left( \frac{1}{2},0,-1 \right) , \nonumber\\
	Q_R: \left( 0,\frac{1}{2},\frac{1}{3} \right),
	\quad \psi_R: \left(0,\frac{1}{2},-1 \right).
\end{eqnarray}
The Higgs sector contains a bidoublet
\begin{equation}
	\phi
	=\begin{pmatrix}
		\phi_1^0 & \phi_1^+ \cr \phi_2^- & \phi_2^0 \cr
	\end{pmatrix}
	:\left(\frac{1}{2}, \frac{1}{2}^\ast, 0\right),
\end{equation}
and two scalar triplets $\Delta_{L,R}$ needed to break the right-handed symmetry,
\begin{eqnarray}
	\Delta_{L,R} = 
	\begin{pmatrix}
		\frac{\Delta_{L,R}^+}{\sqrt{2}}  & 
		\Delta_{_{L,R}}^{^{++}} \cr 
		\Delta_{_{L,R}}^{^0} & 
		\frac{-\Delta_{L,R}^+}{\sqrt{2}}
	\end{pmatrix},
\end{eqnarray}
with the quantum numbers $\Delta_L: (1,0,2) $ and $ \Delta_R: (0,1,2)$, respectively. This choice is by no means unique and there are other multiplet choices that ensure a satisfactory symmetry breaking pattern (we will comment on possible alternatives in the conclusion). For our analysis we adopt, for concreteness, this more standard choice.
The symmetry breaking is triggered by the following vacuum expectation values (VEVs)
\begin{equation}
	\langle \phi \rangle  = 
	\begin{pmatrix}
		\frac{k_1}{\sqrt{2}} & 0                    \\ 
		0                    & \frac{k_2}{\sqrt{2}}
	\end{pmatrix},
	\quad \langle \Delta_{L,R} \rangle = 
	\begin{pmatrix}
		0                        & 0 \\ 
		\frac{v_{L,R}}{\sqrt{2}} & 0
	\end{pmatrix},
\end{equation}
where $v_R$ of the right triplet breaks SU(2)$_R~\otimes$ U(1)$_{B-L}$ to U(1)$_Y$ and provides masses for the new heavy particles. As these new fields have not been observed and due to the strong limits on right-handed currents, $v_R$ should be sufficiently large. On the other hand, the VEV $v_L$ of the left triplet contributes to the $\rho$ parameter, and is therefore experimentally constrained to values $\lesssim 5$~GeV (see for example \cite{delAguila:2011gr}, and references therein). Finally, the bidoublet VEVs $k_1$ and $k_2$ break the Standard Model symmetry and are of the order of the electroweak scale. Consequently, the VEVs follow the hierarchy $ |v_L| \ll |k_i| \ll |v_R|$. In the following we assume that all VEVs are real and we neglect the possibility of $CP$ violating phases.

The leptonic Yukawa Lagrangian under \lr is given by~\cite{Barenboim:1996vu}
\begin{equation}
\label{eq:Lagrangian}
  - \mathcal{L} = 
    \bar\psi_L Y_1 \phi \psi_R 
  + \bar\psi_L Y_2 \tilde\phi \psi_R
  + \psi_L^T (i Y_M) \Delta_L \psi_L
  + \psi_R^T (i Y_M) \Delta_R \psi_R + \text{h.c.},
\end{equation}
where we suppress gauge invariant field contractions and the summation over the flavour indices of the fermion fields and the $3\times 3$ Yukawa matrices $Y_1$, $Y_2$ and $Y_M$.

The Lagrangian (\ref{eq:Lagrangian}) leads to the $6\times 6$ neutrino mass matrix $M_\nu$ with Dirac mass terms arising from $Y_1$ and $Y_2$ as well as Majorana mass terms originating from $Y_M$. The mass matrix can be written in block form
\begin{equation}
\label{eq:matr}
	M_\nu =
	\begin{pmatrix}
		M_L & M_D \\ 
		M_D^T & M_R
	\end{pmatrix},
\end{equation}
in the basis $(\nu_L, \nu^c_L)^T$. Due to the Pauli principle, $M_\nu$ is complex symmetric. The $3\times 3$ entries of this matrix are given by 
\begin{equation}
\label{eq:sub}
	M_{L,R} = \sqrt{2}           Y_M v_{L,R},        \quad 
	M_D     = \frac{1}{\sqrt{2}}(Y_1 k_1 + Y_2 k_2). \quad
\end{equation}
Assuming that all Yukawa couplings (of a given generation) are of similar magnitude, the structure of $M_\nu$ follows the hierarchy of the Higgs VEVs, i.e. $M_L \ll M_D \ll M_R$. Equivalently, the Yukawa couplings in (\ref{eq:Lagrangian}) lead to the charged lepton mass matrix $M_\ell= \frac{1}{\sqrt{2}} (Y_2 k_1 + Y_1 k_2)$. In the following, we will not specify the flavour structure generated by the Yukawa couplings. Instead we will phenomenologically parametrize the charged current couplings between the heavy neutrinos and the charged leptons.  

The neutrino mass matrix is diagonalized by the $6\times 6$ complex orthogonal mixing matrix $U_\nu$ as~\cite{schechter:1980gr}
\begin{equation}
\label{eq:diag}
	U_\nu^T M_\nu U_\nu = 
	\text{diag}(m_{\nu_1}, m_{\nu_2}, m_{\nu_3}, m_{N_1}, m_{N_2}, m_{N_3}),
\end{equation}
with the light and heavy neutrino masses $m_{\nu_i}$ and $m_{N_i}$, respectively (the transformation can be chosen such that the mass eigenvalues are positive and appropriately ordered).

Before discussing the generic flavour structure of the charged currents generated in the LRSM, we will briefly summarize the spectrum of heavy particles relevant to our discussion. After both LR and electroweak symmetry breaking and assuming CP invariance, the charged gauge boson eigenstates $W_L$ and $W_R$ mix as
\begin{equation}
	\begin{pmatrix}
		W_1 \\ W_2
	\end{pmatrix} = 
	\begin{pmatrix}
		 \cos\zeta_W & \sin\zeta_W \\
		-\sin\zeta_W & \cos\zeta_W
	\end{pmatrix}
	\begin{pmatrix}
		W_L \\ W_R
	\end{pmatrix}.
\end{equation}
Similarly, the neutral gauge bosons $W^3_L$, $W^3_R$ and $Y$ combining to $Z_L$ and $Z_R$ mix to form the mass eigenstates
\begin{equation}
	\begin{pmatrix}
		Z_1 \\ 
		Z_2
	\end{pmatrix} = 
	\begin{pmatrix}
		 \cos\zeta_Z & \sin\zeta_Z \\
		-\sin\zeta_Z & \cos\zeta_Z
	\end{pmatrix}
	\begin{pmatrix}
		Z_L \\ Z_R
	\end{pmatrix}.
\end{equation}
As outlined in Section~\ref{sec:limits}, the mixing angles $\zeta_W$ and $\zeta_Z$ are highly constrained from electroweak precision data. Consequently, the mass eigenstates are essentially given by the gauge eigenstates, and the heavy components have masses of the order of the right-handed symmetry breaking scale, $m_{Z_R} \approx m_{W_R} \approx v_R$.

Generally, the neutrino mass eigenstates are mixtures of the weak eigenstates, forming three light neutrinos $\nu_i$ and three heavy neutrinos $N_i$, cf. Eq.~(\ref{eq:diag}). Neglecting the flavour structure, this leads to the diagonalization in block form
\begin{equation}
	\begin{pmatrix}
		\nu \\ 
		N
	\end{pmatrix} = 
	\begin{pmatrix}
		 \cos\zeta_\nu & \sin\zeta_\nu \\
		-\sin\zeta_\nu & \cos\zeta_\nu
	\end{pmatrix}
	\begin{pmatrix}
		\nu_L + \nu_L^C \\ 
		\nu_R + \nu_R^C
	\end{pmatrix},
\end{equation}
with the mixing angle $\zeta_\nu$ between left- and right-handed neutrinos. For one generation, the charged current weak interactions can then be written in terms of the mass eigenstates as
\begin{align}
\label{eq:LRCC0}
	J_{W_1}^{\mu-} &=
		  \frac{g_L}{2\sqrt{2}} \left(\bar\nu + \sin\zeta_\nu\bar N^{c}\right) 		
		  \gamma^\mu(1-\gamma_5) e
		+ \frac{g_R}{2\sqrt{2}} \sin\zeta_W \bar N \gamma^\mu(1+\gamma_5)e,     \nonumber\\
	J_{W_2}^{\mu-} &= 
		- \frac{g_L}{2\sqrt{2}} \sin\zeta_W \bar\nu\gamma^\mu (1-\gamma_5)e 
		+ \frac{g_R}{2\sqrt{2}} \left(\bar N-\sin\zeta_\nu\bar\nu^c \right)
		  \gamma^\mu (1+\gamma_5)e.
\end{align}

\subsection{Generic Lepton Flavour Structure}
\label{sec:FlavourStructure}

In order to discuss the flavour structure of the charged current interaction we need to consider the multi-generation case. 
The effective $3\times 6$ lepton mixing matrices $(U_{L,R})_{\ell i}$, $\ell=e, \mu, \tau$, $i=1,\dots,6$ characterizing the charged (and neutral) current weak interactions of the mass eigenstate neutrinos in any seesaw model have been fully characterized in Ref.~\cite{schechter:1980gr} and may be written as
\begin{equation}
	(U_L)_{\ell i} = \sum_{n=1}^3 \Omega^*_{n \ell} (U_\nu)_{ni},    \quad
	(U_R)_{\ell i} = \sum_{n=1}^3 \Omega^*_{n \ell} (U_\nu)_{n+3,i},
\end{equation}
where $\Omega$ is the $3\times 3$ unitary matrix that diagonalizes the charged lepton mass matrix $M_\ell$, while $U_\nu$ is the $6\times 6$ unitary matrix that diagonalizes the neutrino mass matrix defined in Eq.~(\ref{eq:diag}).
For our purposes we take the charged lepton mass matrix in diagonal form\footnote{This may be automatic in the presence of suitable discrete flavour symmetries as in \cite{Hirsch:2009mx}.} so that $\Omega \to 1$. In this case the overall neutrino matrix $U_\nu$ can be decomposed as
\begin{equation}
	U_\nu = 
	\begin{pmatrix}
		U_L^* \\
		U_R
	\end{pmatrix} =
	\begin{pmatrix}
		U^{LL} & U^{LR} \\
		U^{RL} & U^{RR}
	\end{pmatrix},
\end{equation}
where $U_L$ and $U_R$ relate the left-handed and right-handed neutrino flavour eigenstates $\nu_{L_\ell} = \nu_\ell$ and $\nu_{R_\ell} = N_\ell$ with the mass eigenstates $\nu_i$,
\begin{equation}
	\nu_{{L,R}_\ell} = (U_{L,R})_{\ell i} \nu_i, \quad i=1,\dots,6, 
\end{equation}
and the pieces $U^{LL}$, $U^{LR}$, $U^{RL}$ and $U^{RR}$ can be calculated numerically or may be obtained in seesaw perturbation theory~\cite{schechter:1981cv}.

The charged weak interactions of the light mass eigenstate neutrinos are effectively described by the mixing matrix $U^{LL}$ which is non-unitary, hence the coupling of a given light neutrino to the corresponding charged lepton is decreased with respect to that of the Standard Model. This affects the rates for {\it low energy} weak decay processes, where the states that can be kinematically produced are only the light neutrinos. Similarly, right-handed neutrinos would be produced singly in the decays of the $Z$ at the CERN LEP Collider~\cite{dittmar:1989yg}. As a result there are constraints on the strength of the $U^{LR} \sim \sin\zeta_\nu$ mixing matrix elements that follow from all these measurements.

This formalism can be easily adapted to any seesaw model, such as the \lr scheme described in Section~\ref{sec:minimalLR}. The charged current weak interactions in the left-right model can be written in terms of the neutrino mass eigenstates as,
\begin{align}
\label{eq:LRCC1}
	J_{W_1}^{\mu-} & = 
		  \frac{g_L}{\sqrt{2}} 
		  \left( \bar\nu_i U^{LL}_{\ell i} + \bar N_i^c U^{LR}_{\ell i} \right) 
		  \gamma^\mu \ell_L
		+ \frac{g_R}{\sqrt{2}}\sin\zeta_W
        \left( \bar\nu_i U^{RL}_{\ell i} + \bar N_i   U^{RR}_{\ell i} \right)
        \gamma^\mu \ell_R, \nonumber\\
	J_{W_2}^{\mu-} & =
		- \frac{g_L}{\sqrt{2}}\sin\zeta_W 
        \left( \bar\nu_i U^{LL}_{\ell i} + \bar N_i    U^{LR}_{\ell i} \right) 
        \gamma^\mu \ell_L 
		+ \frac{g_R}{\sqrt{2}} 
		  \left( \bar N_i  U^{RR}_{\ell i} + \bar\nu_i^c U^{RL}_{\ell i} \right)
		  \gamma^\mu \ell_R,
\end{align}
in analogy with Eq.~(\ref{eq:LRCC0}). Here $U^{LL}, U^{RR}$=$\cal O$(1) while the ``small'' terms are $U^{LR}, U^{RL} \sim M_D M_R^{-1} \sim \sin\zeta_\nu$, all of which can be obtained within the seesaw perturbative diagonalization method developed in Ref.~\cite{schechter:1981cv}. In the following, we will assume the limit in which all left-right mixing terms can be neglected, $\sin\zeta_W, \sin\zeta_Z, \sin\zeta_\nu \ll 1$. The applicability of this approximation for our calculations will be quantified below. As a result, the only terms surviving in (\ref{eq:LRCC1}) are 
\begin{align}
	J_{W_L}^{\mu-} &\approx 
	\frac{g_L}{\sqrt{2}} U_{\ell i} \bar\nu_i \gamma^\mu \ell_L,       \nonumber\\ 	
	J_{W_R}^{\mu-} &\approx 
	\frac{g_R}{\sqrt{2}} V_{\ell i} \bar N_i  \gamma^\mu \ell_R,
\end{align}
where we identified $U \equiv U_\text{PMNS} \equiv U^{LL}$ and $V \equiv U^{RR}$ for notational simplicity. The first term describes the mixing of light neutrinos in charged current interactions giving rise to neutrino oscillations, whereas the second term is responsible for the LHC process in Figure~\ref{fig:diagramsLHC}(a) as well as all low energy LFV processes as we will discuss below. We will assume manifest left-right symmetry of the gauge couplings, i.e. $g_R = g_L$, but the general case can always be recovered by simple rescaling.
 
\section{Experimental Data}
\label{sec:limits}

We now turn to the existing experimental constraints on \lr models. Experimental limits on the mass scales and mixing in the minimal LR symmetry model come from a variety of sources. For example relevant constraints can be derived from the $K_L-K_S$ mass difference, $B_d \bar B_d$ oscillations, $b$ quark semileptonic branching ratio and decay rate, neutrinoless double beta decay, universality tests, nonleptonic kaon decays, muon decay, lepton flavour violating processes and astrophysical constraints from nucleosynthesis and SN 1987A \cite{Langacker:1989xa, Zhang:2007da, Maiezza:2010ic}. 
The impact of searches for neutrinoless double beta decay and lepton flavour violating processes will be discussed in more detail in Sections~\ref{sec:limits0nubb} and~\ref{sec:limitsLFV}, respectively. 
Apart from these, the most relevant constraints for our discussions are: A lower bound on the $W_R$ mass of $m_{W_R} > 1.6$~TeV \cite{Beall:1981ze, Barenboim:1996nd} due to CP violating effects stemming from the measurement of the $K_L - K_S$ Kaon mass difference, but with uncertainties from low energy QCD corrections. A more severe limit of $m_{W_R} > 2.5$~TeV was reported in \cite{Zhang:2007da, Maiezza:2010ic}, and the authors of reference \cite{Chakrabortty:2012pp} argue that heavy Higgs masses lighter than 10~TeV are disfavoured in the minimal LRSM due to low energy precision data. The minimal LRSM is therefore strongly constrained by low energy observations. At the Tevatron, searches for $W_R \to e\nu$ yield a limit of $m_{W_R} > 1.12$~TeV at 95\% C.L., assuming SM strength couplings~\cite{Aaltonen:2010jj}. Only recently, these have been superseded by limits derived from searches at the LHC for $W_R$ with decays into $e\nu$ ($m_{W_R} > 1.36$~TeV) \cite{Khachatryan:2010fa} and $\mu\nu$ ($m_{W_R} > 1.40$~TeV) \cite{Chatrchyan:2011dx}, again assuming SM strength couplings. Using data collected in 2010 and 2011 with an integrated luminosity of 240~pb$^{-1}$, the CMS collaboration has reported on the search for the production of $W_R$ bosons and heavy right-handed neutrinos in the minimal LRSM as in Figure~\ref{fig:diagramsLHC}(a). No excess has been observed, thereby excluding a region in the $(m_{W_R}, m_N)$ parameter space extending to $(m_{W_R}, m_N) \approx (1.7\text{ TeV}, 0.6\text{ TeV})$~\cite{CMS:925203}. Even more recently, the ATLAS collaboration has reported on a search for heavy neutrinos and right-handed $W$ bosons via the same channel with an integrated luminosity of 2.1~fb$^{-1}$~\cite{CERN-PH-EP-2012-022}. Again, no excess has been found in this search as well, resulting in an excluded parameter region extending to $(m_{W_R}, m_N) \approx (2.5\text{ TeV}, 1.5\text{ TeV})$. If lighter than $\approx 5$~GeV, right-handed neutrinos can be produced on-shell in $B$ meson decays, and searches at LHCb provide limits on their coupling strength to muons~\cite{Aaij:2011ex, Aaij:2012zr}. The bound on the mixing angle between $W_R$ and $W_L$ is of the order $\zeta_W < \mathcal{O}(10^{-2})$ \cite{Langacker:1989xa, nakamura2010review}.

Direct limits on the $Z_R$ mass from electroweak precision data, such as lepton universality at the $Z$ peak, are of the order $\mathcal{O}(1)$~TeV \cite{Polak:1991pc, Pilaftsis:1995tf, Czakon:1999ga}. Within the minimal LRSM one also has the theoretical relation $m_{Z_R} \approx 1.7 m_{W_R}$ for $g_R \approx g_L$, so that indirect limits via the bounds on the $W_R$ mass also yield more stringent constraints on the $Z_R$ mass. From the same data, the mixing angle between $Z_R$ and $Z_L$ is constrained to be $\zeta_Z < \mathcal{O}(10^{-4})$.

\subsection{Light Neutrino Oscillation Data}
\label{sec:limitsOscillation}

When combined with reactor and accelerator results, solar and atmospheric neutrino experiments~\cite{fukuda:1998mi, ahmad:2002jz, eguchi:2002dm} provide firm evidence for neutrino oscillations~\cite{Maltoni:2004ei}.
Together with reactor data, the experimental results on solar neutrinos clearly suggest $\nu_e \to \nu_{\mu,\tau}$ oscillations driven by the mass squared difference $\Delta m^2_{12} = m_2^2 - m_1^2$ in the range of the large mixing angle solution, while the results on atmospheric neutrinos are interpreted by $\nu_{\mu }\to \nu_{\tau}$ oscillations driven by $\Delta m^2_{23} = m_3^2 - m_2^2$ are characterized by a nearly maximal mixing.  The present global analysis in a three-neutrino framework taking into account the latest results on reactor neutrino fluxes gives the following best fit values for the light neutrino squared mass differences and the mixing angles $\theta_{12}$, $\theta_{23}$ and $\theta_{13}$ of the neutrino mixing matrix $U$ in the standard parametrization~\cite{Tortola:2012te},
\begin{align}
\label{eq:status}
	\Delta m^2_{21}   &= (7.62\pm 0.19) 10^{-5}\eVq,
	\nonumber\\
	\Delta m^2_{31}   &=
	\begin{cases}
		+(2.53^{+0.08}_{-0.10}) 10^{-3}\eVq    & \text{NH} \\
		-(2.40^{+0.10}_{-0.07}) 10^{-3}\eVq    & \text{IH,}
	\end{cases}
	\nonumber\\
	\sin^2\theta_{12} &= 0.320^{+0.015}_{-0.017},
	\nonumber\\
	\sin^2\theta_{23} &=
	\begin{cases}
		0.49^{+0.08}_{-0.05}    & \text{NH} \\
		0.53^{+0.05}_{-0.07}    & \text{IH,}
	\end{cases}
	\nonumber\\
	\sin^2\theta_{13} &=
	\begin{cases}
		0.026^{+0.003}_{-0.004}    & \text{NH} \\
		0.027^{+0.003}_{-0.004}    & \text{IH,}
	\end{cases}
\end{align} 
for normal (NH) and inverse hierarchy (IH), respectively.

\subsection{Neutrinoless Double Beta Decay}
\label{sec:limits0nubb}

%
\begin{figure}[t!]
\centering
\subfigure[]{
\includegraphics[clip,width=0.42\textwidth]{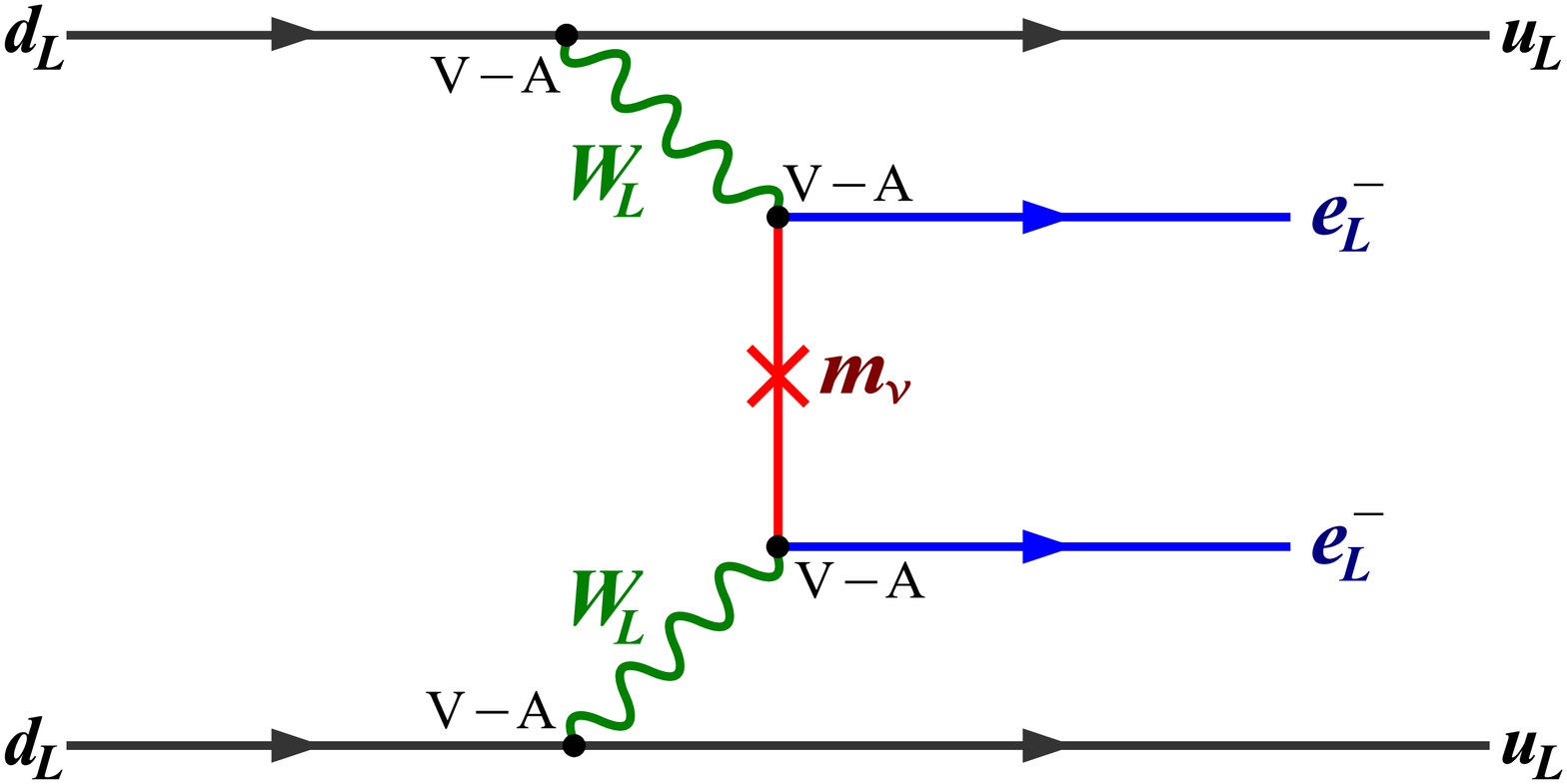}
}
\subfigure[]{
\includegraphics[clip,width=0.42\textwidth]{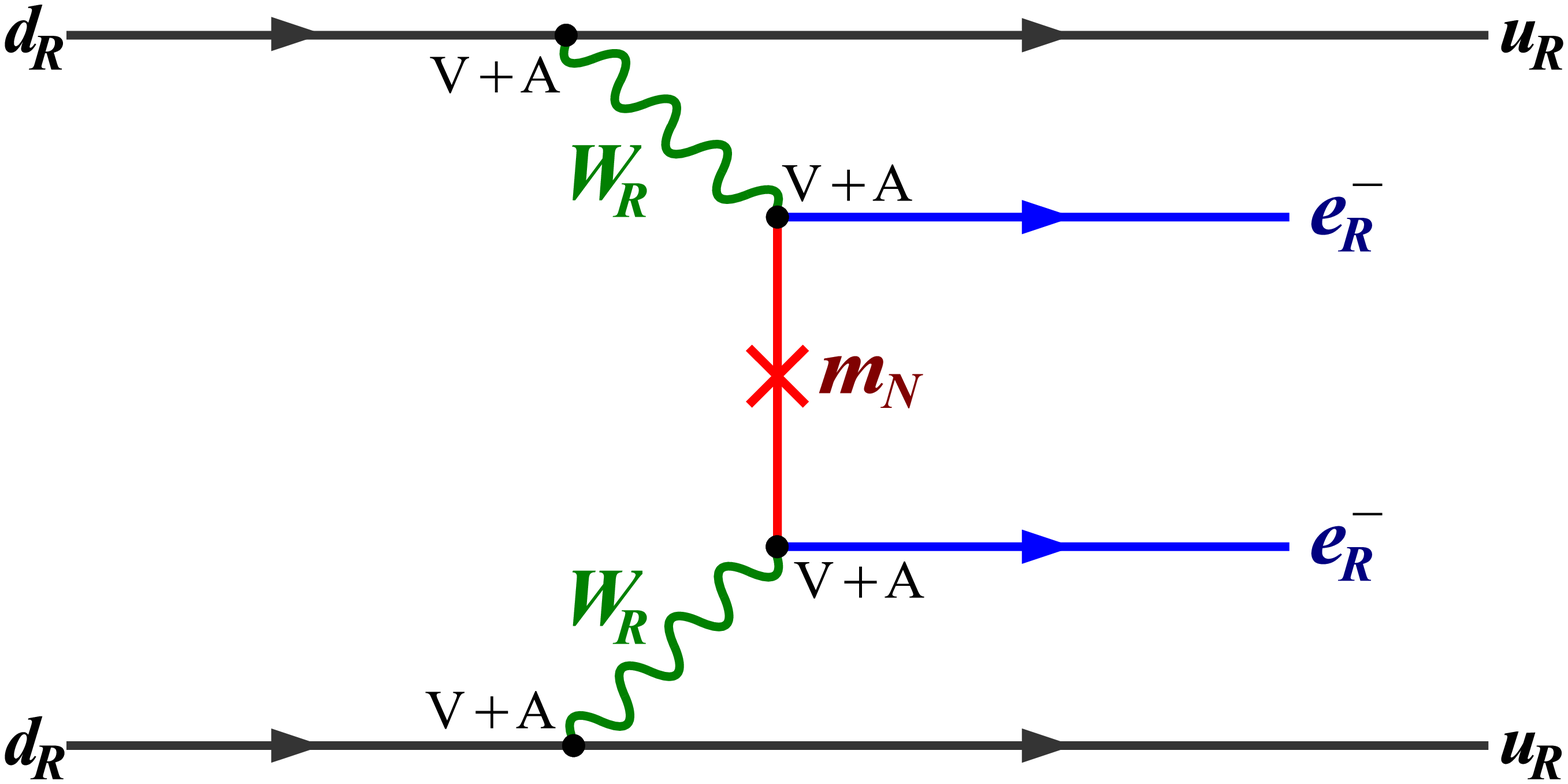}
}\\
\subfigure[]{
\includegraphics[clip,width=0.42\textwidth]{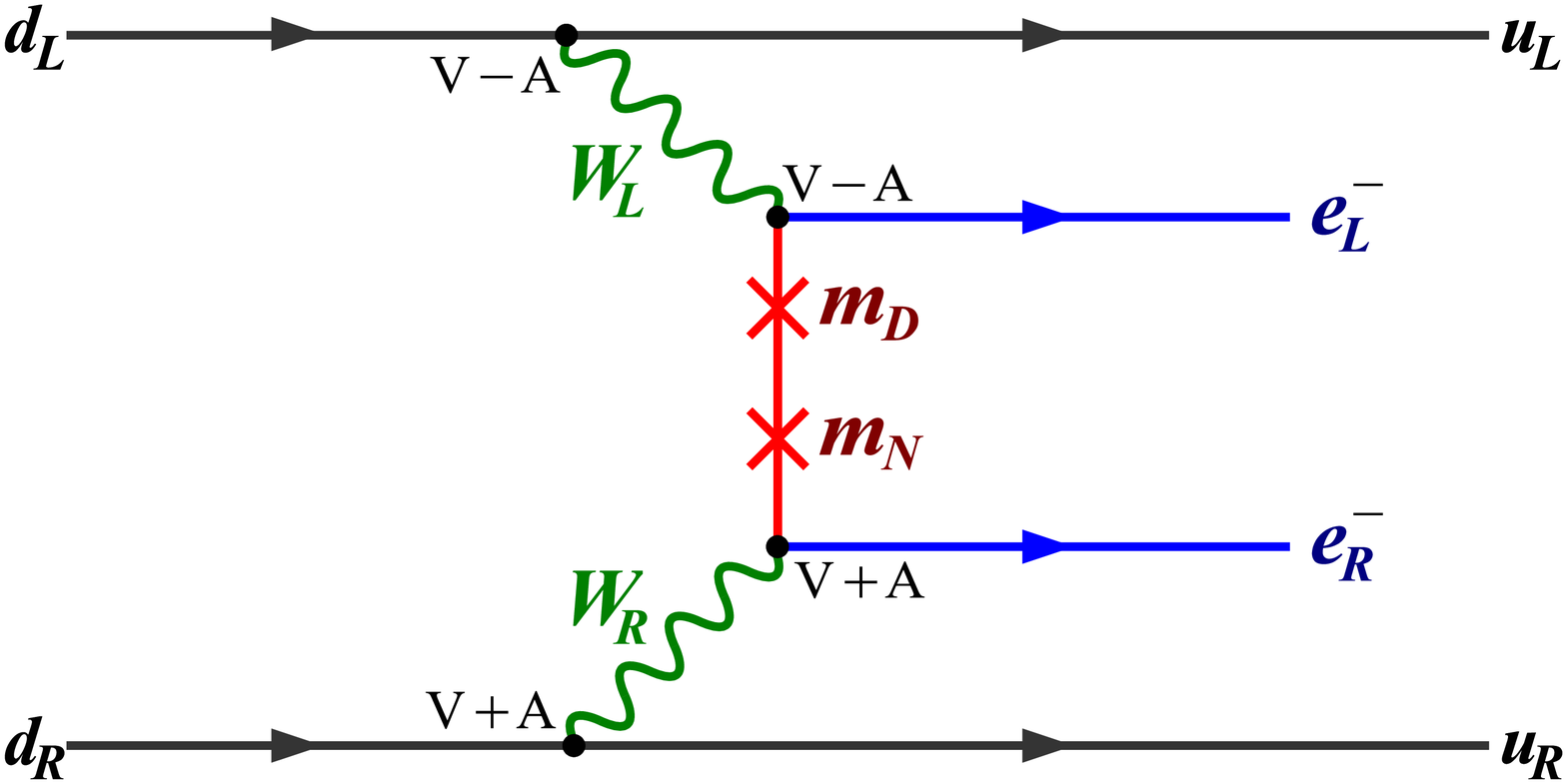}
}
\subfigure[]{
\includegraphics[clip,width=0.42\textwidth]{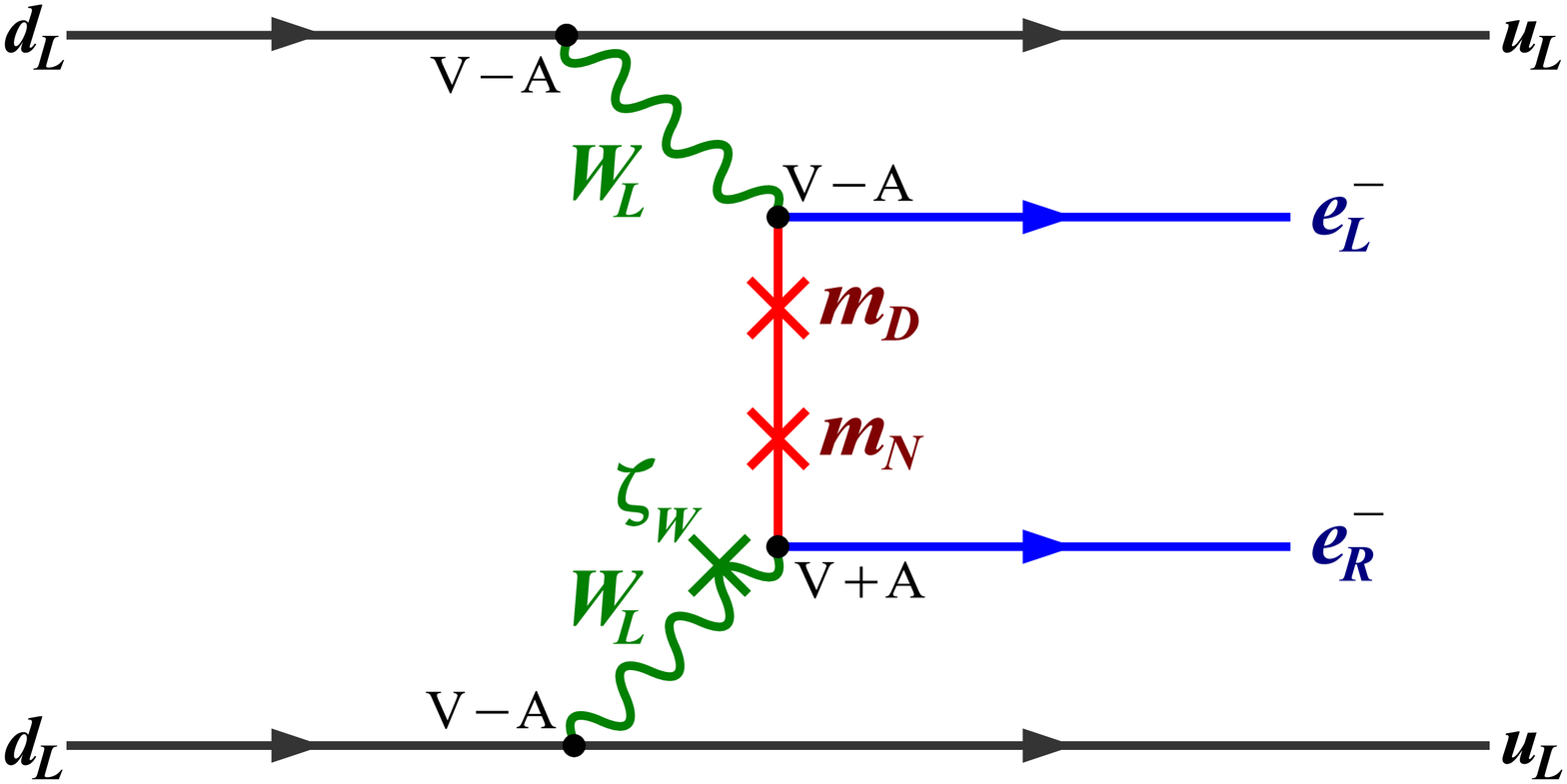}
}\\
\subfigure[]{
\includegraphics[clip,width=0.42\textwidth]{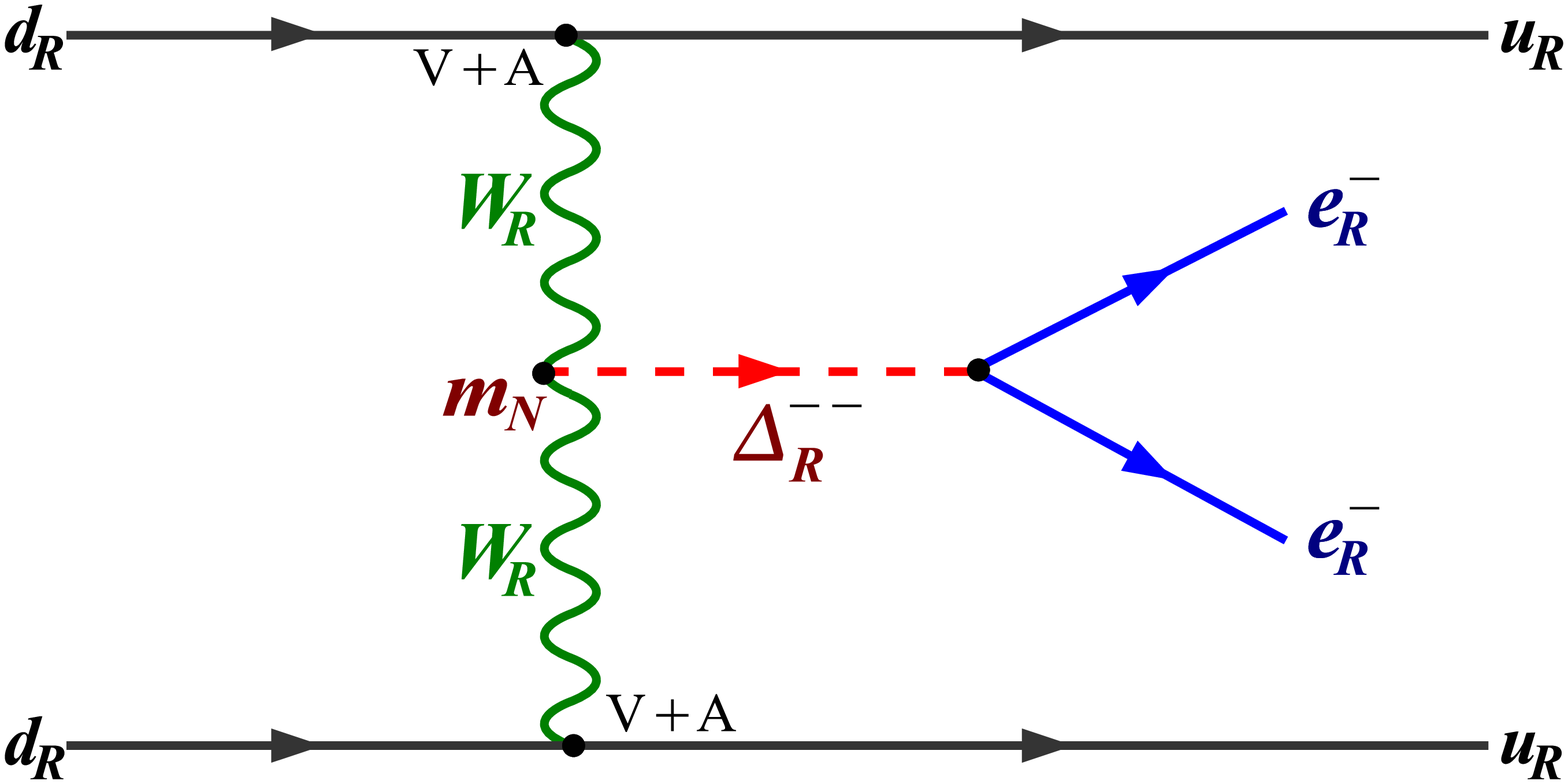}
}
\caption{Diagrams contributing to $0\nu\beta\beta$ decay in left-right symmetry: (a) Light neutrino exchange (standard mass mechanism), (b) Heavy neutrino exchange, (c) Neutrino and heavy $W$ exchange with Dirac mass helicity flip ($\lambda$ mechanism), (d) Neutrino and light $W$ exchange with Dirac mass and $W$ mixing suppression ($\eta$ mechanism), (e) Doubly charged Higgs Triplet exchange.}
\label{fig:diagrams_0nubb} 
\end{figure}
The most sensitive probe of the absolute mass scale of the neutrinos is neutrinoless double $\beta$ decay ($0\nu\beta\beta$). In this process, an atomic nucleus with $Z$ protons decays into a nucleus with $Z+2$ protons and the same mass number $A$ under the emission of two electrons,
\begin{equation}
	\label{eq:0nubb}
	(A,Z)\to(A,Z+2) + 2 e^-.
\end{equation}
This process can be engendered through the exchange of a light neutrino connecting two V-A weak interactions, as illustrated in Figure~\ref{fig:diagrams_0nubb}(a). The process in Eq.~(\ref{eq:0nubb}) is lepton number violating and, in the standard picture of light neutrino exchange, it is only possible if the neutrino is identical to its own anti-particle, i.e. if neutrinos are Majorana particles. In fact quite generally one may argue that, whatever the underlying mechanism inducing $0\nu\beta\beta$~decay, its observation implies the Majorana nature of neutrinos~\cite{schechter:1982bd,Duerr:2011zd}.

Currently, the best limit on $0\nu\beta\beta$~decay comes from the search for $0\nu\beta\beta$~decay of the isotope $^{76}$Ge giving a half-life of $T_{1/2} > 1.9\cdot10^{25}$~yrs~\cite{KlapdorKleingrothaus:2000sn}. This results in an upper bound on the effective $0\nu\beta\beta$ Majorana neutrino mass of $m_{\beta \beta} \equiv |\sum_i U^2_{ei}m_{\nu_i}| < 300-600$~meV, depending on the model used to calculate the nuclear matrix element of the process. A controversial claim of observation of $0\nu\beta\beta$~decay in $^{76}$Ge gives a half-life of $T_{1/2}=(0.8-18.3) \cdot 10^{25}$~yrs \cite{klapdor-kleingrothaus:2001ke} and a resulting effective Majorana neutrino mass of $m_{\beta \beta} = 110-560$~meV. Next generation experiments such as SuperNEMO, GERDA, CUORE, EXO or MAJORANA aim to increase the half-life sensitivity by one order of magnitude and will confirm or exclude the claimed observation. The planned experiment SuperNEMO allows the measurement of $0\nu\beta\beta$~decay in several isotopes to the ground and excited states and is able to track the trajectories of the emitted electrons and determine their individual energies. In this respect, the SuperNEMO experiment has a unique potential to disentangle the possible mechanisms for $0\nu\beta\beta$~decay~\cite{Arnold:2010tu, Deppisch:2010zza}.

In the left-right symmetric model, several such mechanisms can contribute to  $0\nu\beta\beta$ as shown in Figure~\ref{fig:diagrams_0nubb}. Here, contributions (a)-(d) are due to the exchange of either light or heavy neutrinos as well as light and heavy $W$ bosons. All these terms could be described by a single Feynman diagram using the mass eigenstates $n_i$ ($i=1,...,6$) and $W_a$ ($a=1,2$) of neutrinos and $W$ bosons. The separation into the four contributions shown in Figure~\ref{fig:diagrams_0nubb} is traditionally used as it illustrates the dependence on and suppression with the different LRSM model parameters. Diagram~\ref{fig:diagrams_0nubb}(a) describes the exchange of massive light neutrinos corresponding to the generally considered mass mechanism. Its contribution to $0\nu\beta\beta$ depends on the effective neutrino mass	$m_{\beta\beta} = |\sum_i U_{ei}^2 m_{\nu_i}|$, and saturates current experimental bounds if the light neutrinos are degenerate with mass scale $m_{\nu_1} \approx m_{\beta\beta} \approx 0.3 - 0.6$~eV.

Correspondingly, diagram~\ref{fig:diagrams_0nubb}(b) describes the exchange of heavy right-handed neutrinos and depends on the effective coupling
\begin{equation}
\label{eq:epsilonN}
	\epsilon_N = \sum_{i=1}^3 V_{ei}^2 	
	\frac{m_p}{m_{N_i}}\frac{m_{W_L}^4}{m_{W_R}^4}.
\end{equation}
If this is the dominant contribution to $0\nu\beta\beta$, current experimental limits correspond to $|\epsilon_N| \lsim 2 \times 10^{-8}$~\cite{Rodejohann:2011mu}.

Diagrams~\ref{fig:diagrams_0nubb}(c) and (d) are suppressed by the left-right mixing $M_D/M_N \sim \zeta_\nu \sim \sqrt{m_\nu/m_N}$ (the latter relation is valid for a dominant type-I seesaw mass mechanism~\cite{schechter:1981cv}) between light and heavy neutrinos. In our approach we assume that this mixing is small enough so that decays of heavy neutrinos via this Yukawa coupling are negligible compared to the three-body decays via the $SU(2)_R$ gauge coupling. In this case, these contributions to $0\nu\beta\beta$ are also generally negligible, and we will not discuss them further here.

Diagram~\ref{fig:diagrams_0nubb}(e) describes the contribution from the exchange of a right-handed doubly-charged triplet Higgs $\Delta_R^{--}$\footnote{A priori, there is an analogous diagram with a left-handed doubly-charged Higgs, but its contribution is always sub-dominant to the standard mass mechanism unless there is a fine-tuning between type-I and type-II seesaw contributions to the light neutrino masses.}, with the effective coupling
\begin{equation}
\label{eq:epsilonDelta}
	\epsilon_\Delta = \sum_{i=1}^3 V_{ei}^2 		
	\frac{m_{N_i}m_p}{m^2_{\Delta_R^{--}}}\frac{m_{W_L}^4}{m_{W_R}^4}.
\end{equation}
If dominant, current experimental limits correspond to
$|\epsilon_\Delta| \lsim 8 \times 10^{-8}$~\cite{Rodejohann:2011mu}.

\subsection{Low Energy Lepton Flavour Violating Processes}
\label{sec:limitsLFV}

%
\begin{figure}[t!]
\centering
\includegraphics[clip,width=0.50\textwidth]{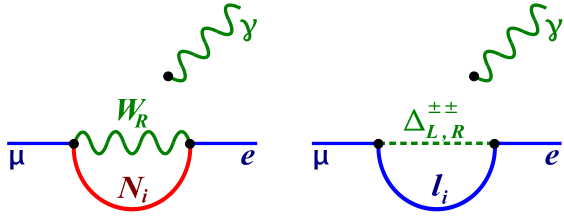}
\caption{Diagrams contributing to $\mu\to e\gamma$. To form complete
  diagrams, the external photon can be attached to any charged
  particle line.}
\label{fig:diagrams_LFV_muegamma} 
\end{figure}
\begin{figure}[t]
\centering
\includegraphics[clip,width=0.49\textwidth]{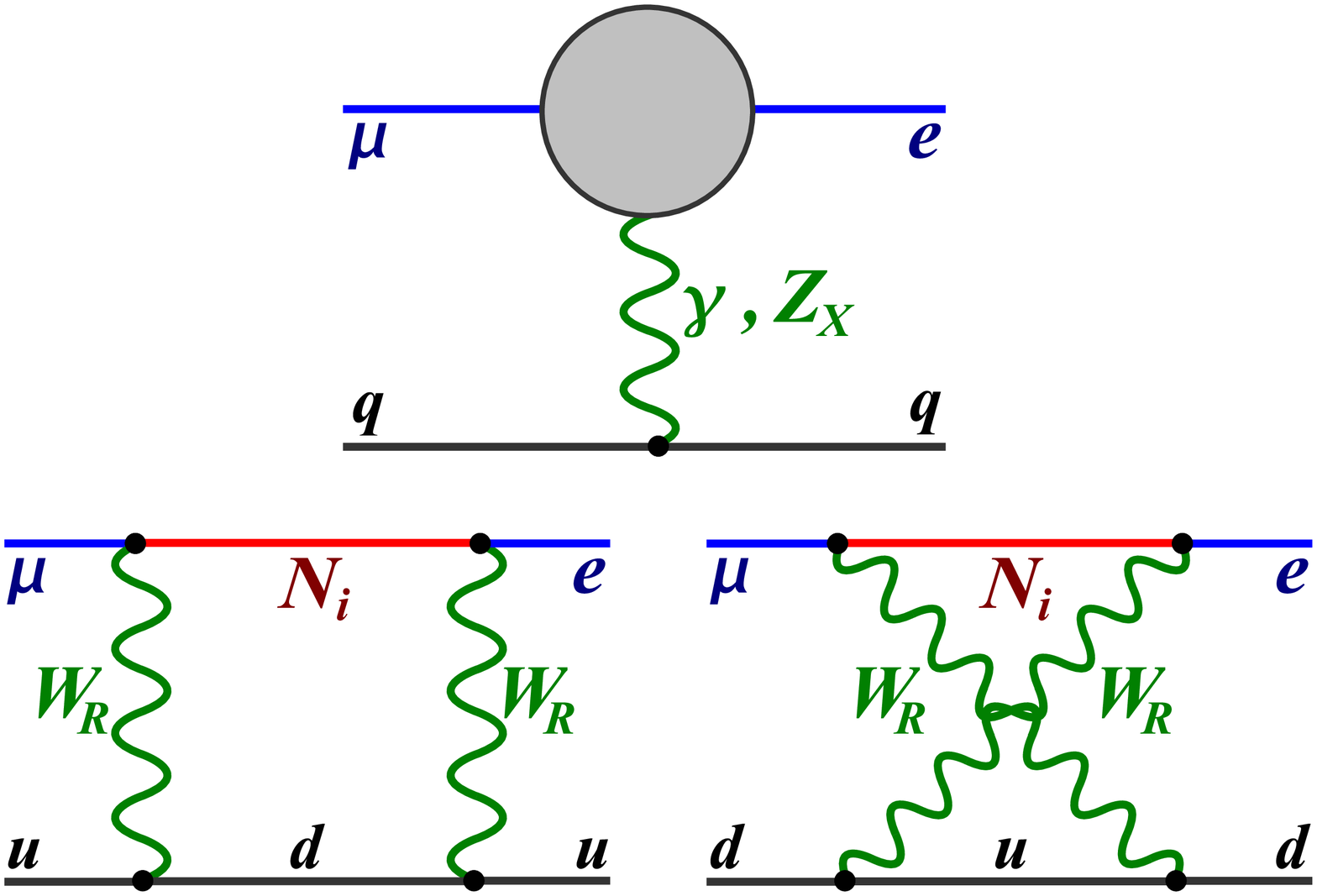}
\includegraphics[clip,width=0.49\textwidth]{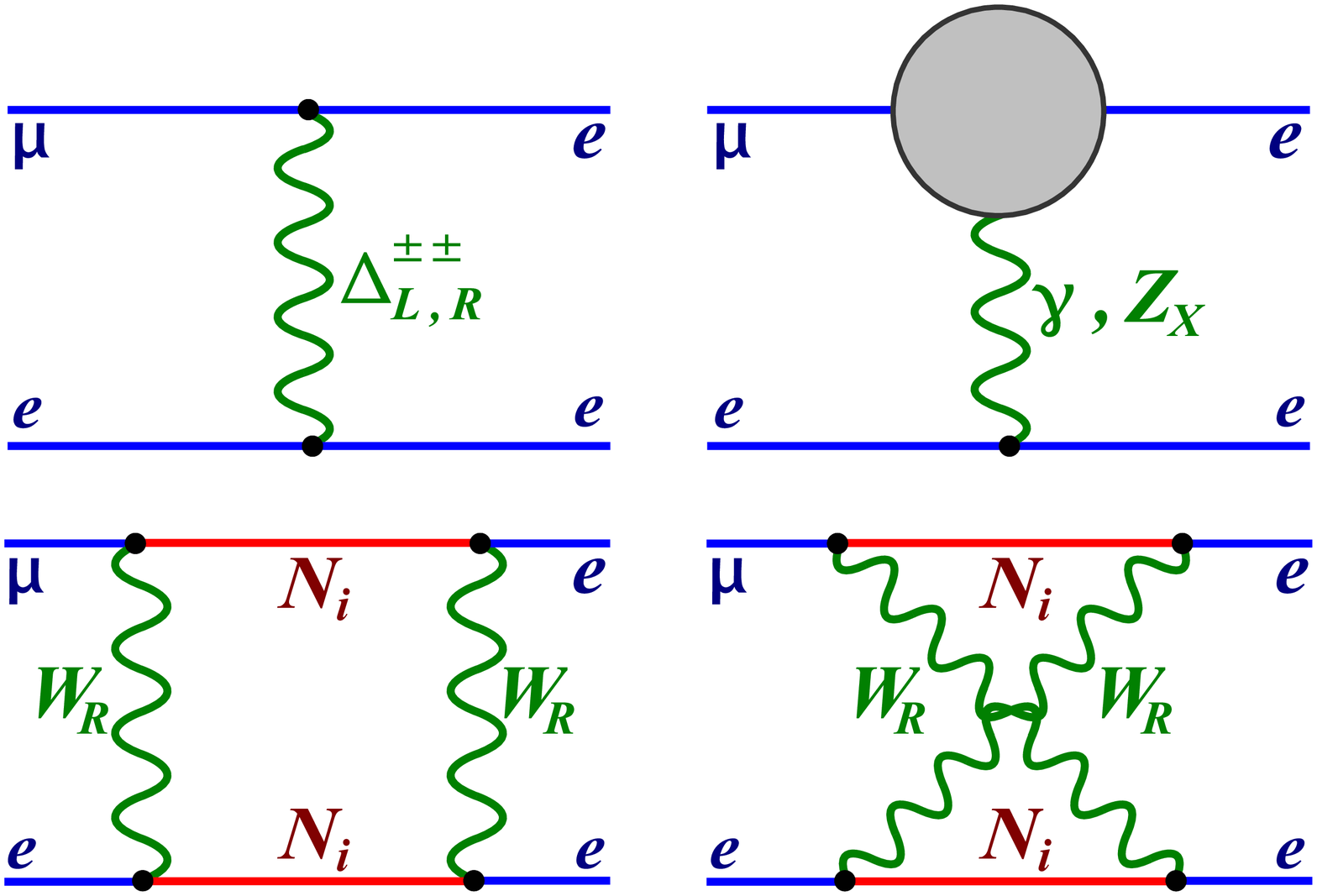}
\caption{Diagrams contributing to $\mu\to e$ conversion in nuclei (left) and $\mu\to eee$ (right) in left-right symmetry. The grey circle represents the effective $\mu-e-$gauge boson vertex with contributions from Figure~\ref{fig:diagrams_LFV_muegamma}.}
\label{fig:diagrams_LFV_mue} 
\end{figure}

The existence of neutrino oscillations suggests that, at some level, lepton flavour violation should also take place in other processes. When taking into account only light neutrinos, LFV is strongly suppressed by $(\Delta m^2_\nu/m_W^2) \approx 10^{-50}$, due to the GIM mechanism. This results in LFV process rates far below any experimental sensitivity which can be safely ignored. Within the LRSM, charged lepton flavour violation naturally occurs due to potentially large flavour violating couplings of the heavy
right-handed neutrinos and Higgs scalars with charged leptons. Amongst a wide range of possible low energy LFV observables, these give rise to observable rates for the processes $\mu\to e\gamma$, $\mu\to eee$ and $\mu\to e$~conversion in nuclei, cf. Figures~\ref{fig:diagrams_LFV_muegamma} and \ref{fig:diagrams_LFV_mue}.

Taking into account contributions from heavy right-handed neutrinos and Higgs scalars, the expected branching ratios and conversion rates of the above processes have been calculated in the LRSM in \cite{Cirigliano:2004mv}. In general, these depend on many parameters, but under the assumption of similar mass scales between the heavy particles in the LRSM, $m_{N_i} \approx m_{W_R} \approx m_{\Delta_L^{--}} \approx m_{\Delta_R^{--}}$ one can make simple approximations. Such a spectrum is naturally expected, as all masses are generated in the breaking of the right-handed symmetry. Under this assumption, the expected branching ratios are given by~\cite{Cirigliano:2004mv}
\begin{align}
\label{eq:BrmuegammaSimplified}
	Br(\mu\to e\gamma) &\equiv 
	\frac{\Gamma(\mu^+ \to e^+\gamma)}{\Gamma(\mu^+ \to e^+\nu\bar\nu)}
	\nonumber\\
	&\approx 1.5 \times 10^{-7} |g_{e\mu}|^2 
	\left(\frac{1\text{ TeV}}{m_{W_R}}\right)^4, \\
\label{eq:BrmueSimplified}
	R^N(\mu\to e) &\equiv 
	\frac{\Gamma(\mu^- + {}^A_Z N \to e^- + {}^A_Z N)}
	{\Gamma(\mu^- + {}^A_Z N \to \nu_\mu + {}^A_{Z-1} N')} \nonumber\\
	&\approx X_N \times 10^{-7} |g_{e\mu}|^2 
	\left(\frac{1\text{ TeV}}{m_{\Delta_R^{--}}}\right)^4 
	\alpha \left(\log\frac{m^2_{\Delta_R^{--}}}{m^2_{\mu}}\right)^2, \\
\label{eq:BrmueeeSimplified}
	Br(\mu\to eee) &\equiv 
	\frac{\Gamma(\mu^+ \to e^+e^-e^+)}{\Gamma(\mu^+ \to e^+\nu\bar\nu)}
	\nonumber\\
	&\approx \frac{1}{2}|h_{e\mu}h^*_{ee}|^2 
	\left(\frac{m_{W_L}^4}{m^4_{\Delta_R^{--}}} +
	\frac{m_{W_L}^4}{m^4_{\Delta_L^{--}}}\right).
\end{align}
Here, $X_\text{(Al,Ti,Au)} \approx (0.8,1.3,1.6)$ is a nucleus-dependent factor whereas $g_{e\mu}$ and $h_{ij}$ describe the effective lepton-gauge boson couplings and lepton-Higgs coupling in (quasi-)\-manifest left-right symmetry,
\begin{align}
\label{eq:LFVCouplings}
	g_{e\mu} &= 
		\sum_{n=1}^3 V^*_{en} V^{\phantom{\dagger}}_{\mu n}
		\left(\frac{m_{N_n}}{m_{W_R}}\right)^2, \\
	h_{ij} &= 
		\sum_{n=1}^3 V_{in} V_{jn}
		\left(\frac{m_{N_n}}{m_{W_R}}\right), \quad i,j=e,\mu,\tau.
\end{align}
As shown in \cite{Cirigliano:2004mv}, the above approximations are
valid if the masses generated in breaking the right-handed symmetry are
of the same order with $0.2 \lsim m_i/m_j \lsim 5$ for any pair of
$m_{i,j} = m_{N_n}, m_{W_R}, m_{\Delta_L^{--}}, m_{\Delta_R^{--}}$. In
our approach, we keep $m_{W_R}$ and $m_{N_n}$ as free parameters of
the order of 0.5 - 5~TeV, relevant for LHC searches, and consider an
order of magnitude variation in the heavy Higgs masses of $0.3 <
m_{\Delta_{L,R}^{--}}/m_{W_R} < 3$.

Several properties of the
Eqs.~(\ref{eq:BrmuegammaSimplified})-(\ref{eq:BrmueeeSimplified}) can then
be derived: (i) Both $Br(\mu\to e\gamma)$ and $R^N(\mu\to e)$
are proportional to the LFV factor $|g_{e\mu}|^2$. In addition, as
$\alpha\log((5\text{ TeV})^2/m^2_\mu) = \mathcal{O}(1)$, the ratio of their rates gives $R^N(\mu\to e)/Br(\mu\to e\gamma) = \mathcal{O}(1)$, independent
of the right-handed neutrino mixing matrix $V$ and largely independent
of the heavy particle spectrum. This consequence of the logarithmic
enhancement of the doubly-charged Higgs boson contributions to $\mu\to
e$ conversion is in stark contrast to models where the symmetry
breaking occurs far above the electroweak scale, such as in
supersymmetric seesaw models with low $\tan\beta$. Here, the $Z$ and photon penguin
contributions dominate, and $R^N(\mu\to e)/Br(\mu\to e\gamma)
\propto \alpha \lesssim \mathcal{O}(10^{-2})$. (ii) Unless there are
cancellations, the LFV couplings are $|g_{e\mu}| \approx
|h^*_{ee}h_{e\mu}|$ and therefore $Br(\mu\to eee)/R^N(\mu\to e) =
\mathcal{O}(300)$ (for $m_{\Delta_{L,R}^{--}} \approx$ 1 TeV).
 
The above theoretical predictions are to be compared with the current
experimental upper limits at 90\% C.L.~\cite{Adam:2011ch, Bertl:2006up, Bellgardt:1987du},
\begin{align}
\label{eq:Bexpllgamma}
	Br_{\rm exp}(\mu\to e\gamma) &< 2.4 \cdot 10^{-12}, \nonumber\\
	R^{Au}_{\rm exp}(\mu\to e) &< 8.0 \cdot 10^{-13},   \\
	Br_{\rm exp}(\mu\to eee) &< 1.0 \cdot 10^{-12}.     \nonumber 
\end{align}
With the current experimental limits roughly of the same order, it
follows that the most restrictive parameter bounds in the LRSM are
derived from the limits on $Br(\mu\to eee)$.

As for future developments, the currently running MEG
experiment~\cite{Adam:2011ch} aims for a sensitivity of
\begin{equation}
\label{eq:BexpllgammaMEG}
	Br_{\rm MEG} (\mu\to e\gamma) \approx 10^{-13},
\end{equation}
whereas the COMET and Mu2e experiments both plan to reach a sensitivity of~\cite{Kutschke:2011ux, Kurup:2011zza}
\begin{equation}
\label{eq:RmueCOMET}
	R^{Al}_{\rm COMET} (\mu\to e) \approx 10^{-16}.
\end{equation}
%

\section{Dilepton Signals at the LHC}
\label{sec:ResultsDilepton}

In the following we discuss the LHC potential to discover lepton flavour and lepton number violating dilepton signals from the production of a heavy right-handed neutrino, 
\begin{equation}
\label{eq:Nprod}
	p + p \to W_R^\pm \to \ell_a^\pm + N_R,
\end{equation}
followed by a three-body decay of $N_R$, as shown in Figure~\ref{fig:diagramsLHC}(a),
\begin{equation}
\label{eq:Ndec}
	N_R \to \ell_b^\mp + W_R^\ast \to \ell_b^\mp + 2 j.
\end{equation}
This process is the main production channel for the right handed
neutrinos $N_R$, where the cross section is typically enhanced,
compared to other production mechanisms, via on-shell $Z_R$
production (Figure~\ref{fig:diagramsLHC}(b)) and $W_R$ fusion (Figure~\ref{fig:diagramsLHC}(c))~\cite{Keung:1983uu, Ho:1990dt, Ferrari:2000sp, Gninenko:2006br}.

Since the signal has no missing energy, and since the heavy on-shell $W_R$ subsequently decays into an on-shell $N_R$, backgrounds
can be removed and the signal well identified by the two $W_R$
and $N_R$ resonances in the respective invariant mass spectra, see
Figure~\ref{fig:distromass}. The dilepton signals in left-right
symmetric models have already been
studied~\cite{Ferrari:2000sp, Gninenko:2006br, Nemevsek:2011hz}, and bounds on the $W_R$
and $N_R$ masses have been obtained~\cite{Chatrchyan:2011dx, CERN-PH-EP-2012-022}.

Here we extend the existing analyses and include lepton flavour
violation in the right-handed neutrino sector to assess the LHC potential to probe the right-handed neutrino sector. This allows us not only to
determine the $W_R$ and $N_R$ masses, but also to unravel the flavour
mixing pattern in the heavy $N_R$ sector. These results will be
confronted with low energy probes of rare LFV processes and neutrinoless double beta decay.

\subsection{Event Simulation, Topology and Cut Flow}
\label{sec:simulation}

%
\begin{figure}[t]
\centering
\includegraphics[clip,width=0.6\textwidth]{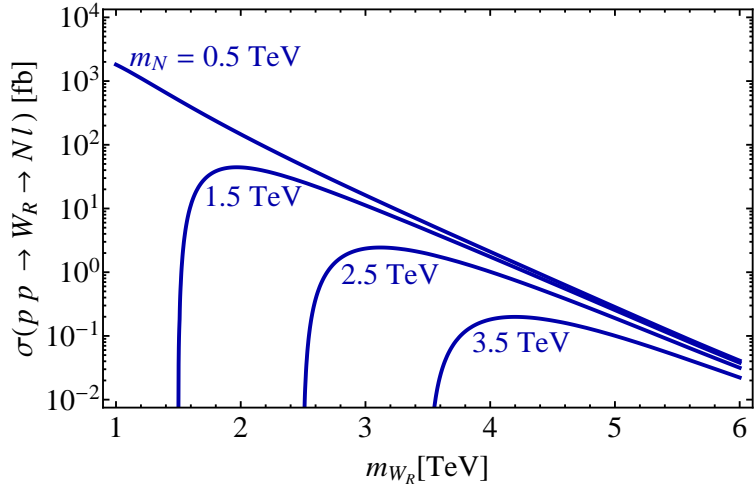}
\caption{Total cross section of the process $pp\to W_R \to N_R \ell$ as a function of the $W_R$ mass, for different values of the neutrino $N_R$ mass, calculated with {\tt PROTOS}~2.1~\cite{protos} for the LHC with $\sqrt s = 14$~TeV.}
\label{fig:crosssection} 
\end{figure}

We generate the partonic signal events $q \bar q \to W_R \to \ell_a
N_R$, followed by the three-body decay $N_R \to
\ell_b jj$,  with the {\tt Triada}~1.1
generator~\cite{delAguila:2008cj} of the Monte Carlo package {\tt PROTOS}~2.1~\cite{protos}. As a benchmark scenario we choose
\begin{equation}
\label{eq:scenario}
	m_{W_R}     =   2~{\rm TeV}, \quad
	m_{N_R}     = 0.5~{\rm TeV},
\end{equation}
where we assume the equality of the two SU(2) gauge couplings for all calculations, $g_R = g_L$, and only one of the three neutrinos 
$N_R$ is lighter than the $W_R$ boson. If other neutrino states are lighter than $W_R$, and if their mass difference is sufficiently large, their signals can be separated by the resonances in the invariant mass distributions, see Figure~\ref{fig:distromass}. Due to QCD radiation, the small inherent widths of the neutrinos of some $10$~keV get broadened up to  $\approx 50$~GeV. As discussed above, we assume negligible left-right mixing, with $\zeta_W$, $\zeta_Z$ and $U^{LR} \sim \zeta_\nu$ smaller than  $10^{-4}$~\cite{delAguila:2009bb}, such that the SM decays $N_R \to W \ell, Z \nu, H \nu$ are sufficiently suppressed\footnote{See Ref.~\cite{delAguila:2009bb} for a discussion of signals with dominating  SM decays of $N_R$.}. For the decays of the heavy $W_R$ boson and the neutrino we obtain the branching ratios $Br(N_R \to \ell \,q q^\prime)\approx 83\%$, for $q=u,c$, $q^\prime= d,s$,  $Br(N_R \to \ell \,t b)\approx 17\%$, and $Br(W_R \to \ell N_R)\approx 9\%$ in our benchmark scenario.

\subsubsection{Signal Topology}

%
\begin{figure}[t]
\centering
\includegraphics[clip,width=0.55\textwidth]{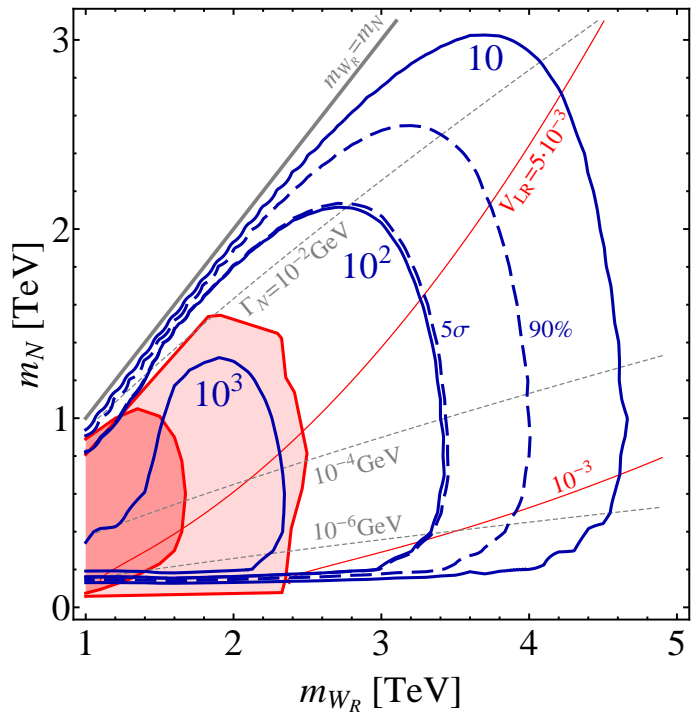}
\caption{Discovery reach of the LHC with $\sqrt{s}=14$~TeV and ${\mathcal L}=30\text{ fb}^{-1}$ in the $m_{W_R}$-$m_N$ parameter plane. The blue solid contours denote constant event rates of the signal process $pp \to W_R \to \ell \ell + 2\text{ jets}$, after event reconstruction and selection cuts (see Section~\ref{sec:reco}). The blue dashed contours indicate the discovery reach with $S/\sqrt{B}=5$, and the exclusion region at the $90\%$ C.L. The grey solid contour indicates the kinematical threshold $m_{W_R} = m_N$ for the on-shell decay $W_R \to \ell N_R$. The grey dashed lines are contours of constant neutrino width $\Gamma_N$.  Above the red solid contours the neutrino decays via an off-shell $W_R$ start to dominate, i.e. $Br(N\to \ell W_R^\ast) > 0.5$, over the SM decay modes $N\to \ell W, \nu Z, \nu H$~\cite{delAguila:2009bb} for a given value of $V^2_{LR} = \sin^2\zeta_W +\sin^2\zeta_\nu$, see Eq.~(\ref{eq:LRCC0}). The red shaded areas are excluded by current LHC searches at CMS~\cite{CMS:925203} (dark shaded) and ATLAS~\cite{CERN-PH-EP-2012-022}~(light shaded).}
\label{fig:MW_MN_ee} 
\end{figure}

In Figure~\ref{fig:crosssection}, we show the $m_{W_R}$ dependence of
the leading order total production cross section $\sigma(pp\to W_R \to
\ell\ell + 2j)$. For $m_{W_R} \approx 2.5$~TeV and $m_N \lesssim 1.5$~TeV (the exclusion reach of current LHC searches), cross sections up to 30 - 50~fb are possible. In Figure~\ref{fig:MW_MN_ee}, we show the expected event rates after reconstruction and cuts (as described below) in the $m_{W_R}$-$m_{N_R}$ plane. To understand the main kinematic features in the production and decay of $W_R$ and $N_R$, the parameter region can be divided into three different kinematic areas:

1) The \emph{threshold region}, with $m_{N_R} \lsim m_{W_R}$, is
close to where the two-body decay $W_R \to \ell N_R$ is kinematically forbidden, and the signal cross section and thus the LHC sensitivity are therefore suppressed.

2) The \emph{jet region}, with $m_{N_R} \ll m_{W_R}$. Due to the large
mass difference, the $N_R$ is highly boosted. The lepton from the
subsequent decay $N_R \to \ell + 2j$ is not isolated but tends to lie inside the cone of the two jets. Although the bare signal cross section
is almost constant for fixed $m_{W_R}$, the LHC sensitivity in this region is quickly suppressed for $m_N \lesssim 200$~GeV.

3) The \emph{discovery reach zone}, with $m_{N_R} < m_{W_R}$ enables ideal
decay kinematics with a large production cross section and isolated leptons such that the best bounds on the ${W_R}$ mass can be set.

The blue contours in Figure~\ref{fig:MW_MN_ee} give an overview of the
LHC reach to probe the $W_R$ and $m_{N_R}$ masses.  As shown by the
red contours in Figure~\ref{fig:MW_MN_ee}, the value of the mixing
parameter $V_{LR}^2=\sin^2\zeta_W +\sin^2\zeta_\nu$, see
Eq.~(\ref{eq:LRCC0}) with $g_L=g_R$, must be chosen sufficiently
small, $V_{LR}\lsim 10^{-3}$ to suppress the SM decays $N\to \ell W,
\nu Z, \nu H$ in the relevant parameter space, which would otherwise lead to trilepton signatures, see Ref.~\cite{delAguila:2009bb}.

\subsubsection{Background Events}

Since our signal consists of two isolated leptons and at least two high $p_T$
jets, the major SM background sources for opposite-sign dileptons stem
from $Z + j (j)$ (one or two hard jets) and $t\bar t$ production. These backgrounds have total cross sections of the order of $2\times 10^3$~pb ($Z + j (j)$) and $5\times 10^2$~pb ($t\bar t$), compared to the maximally possible signal cross section of $\approx 5\times 10^{-2}$~pb. Other SM backgrounds with subdominant cross sections arise from diboson production, $pp \to WW,WZ,ZZ,WH, ZH$. Note that these SM background sources only provide opposite-sign dileptons, whereas same-sign dilepton pairs may originate from charge mis-identification, mistakenly reconstructed leptons from jets and SM diboson production~\cite{Aad:2011vj}, which is generally small compared to opposite-sign background. We use the code {\tt Alpgen}~\cite{Mangano:2002ea} to generate the SM $Z+$jets background, requiring a minimal jet transverse momentum of $p_T>20$~GeV to reduce CPU time, and later using proper matching to take into account final state radiation. The parton shower Monte Carlo {\tt Pythia} 6.4~\cite{Sjostrand:2006za} is used to generate the SM backgrounds $pp \to t\bar t$ at leading order. For $t\bar t$, we take NLO effects into account by
multiplication with a factor $\kappa=1.6$. All hard partonic events are then passed to {\tt Pythia}, to add initial and final state radiation and pile-up, and to perform hadronization. Proper matching is applied for initial and final state radiation for the $Z$+jets {\tt Alpgen} events.

\begin{figure}[t]
\centering
\subfigure[]{
\includegraphics[clip,width=0.448\textwidth]{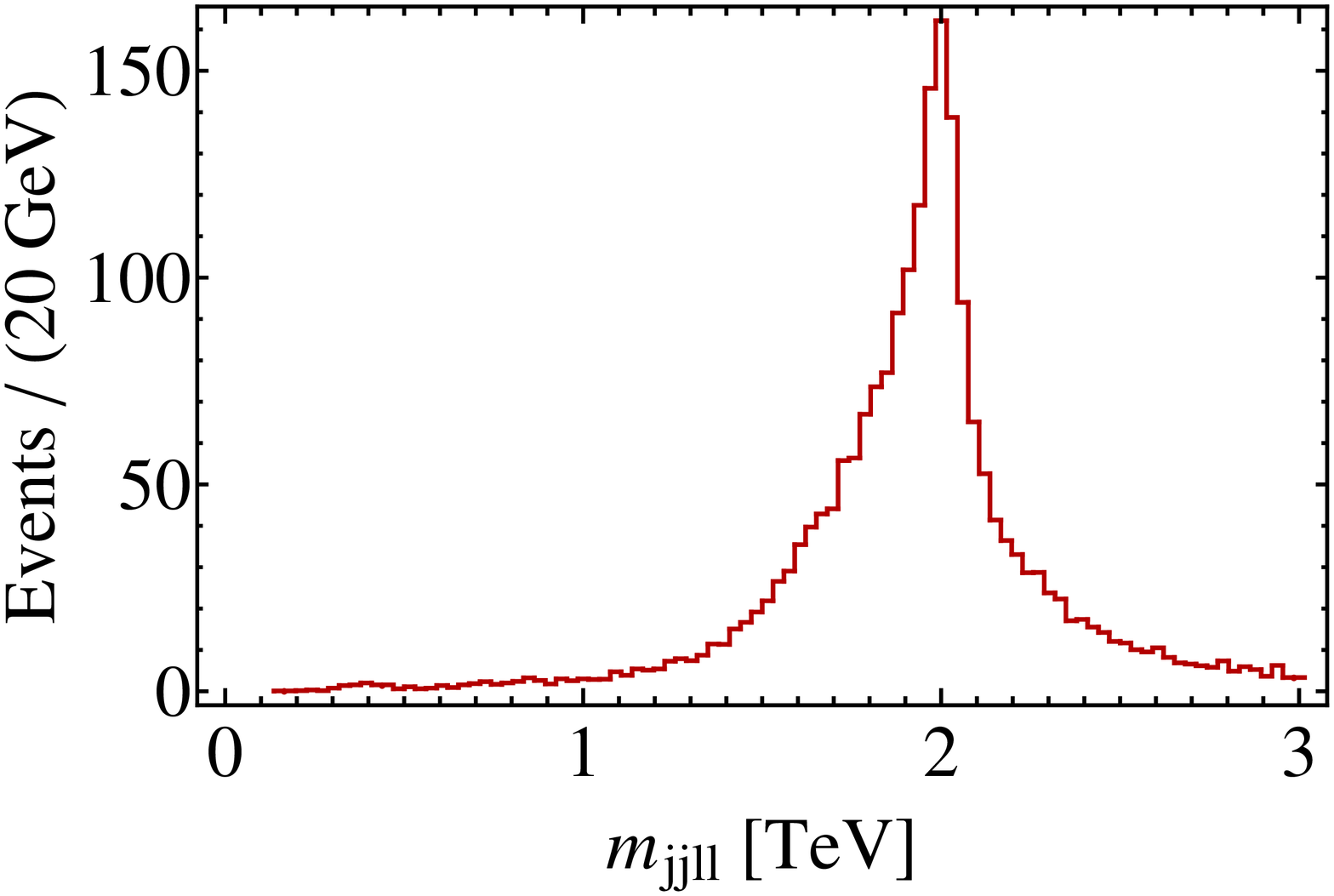}
}
\subfigure[]{
\includegraphics[clip,width=0.45\textwidth]{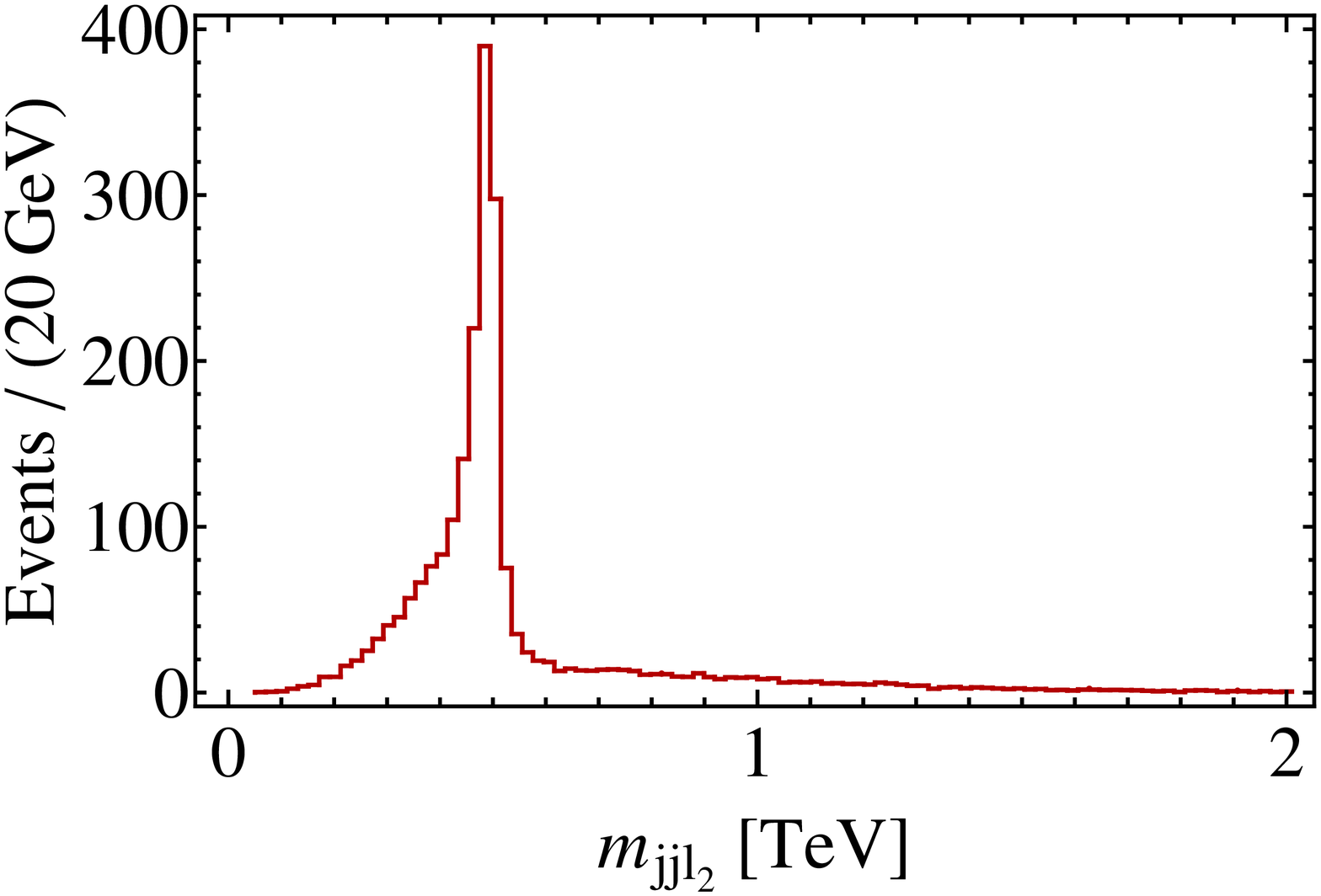}
}\\
\subfigure[]{
\includegraphics[clip,width=0.448\textwidth]{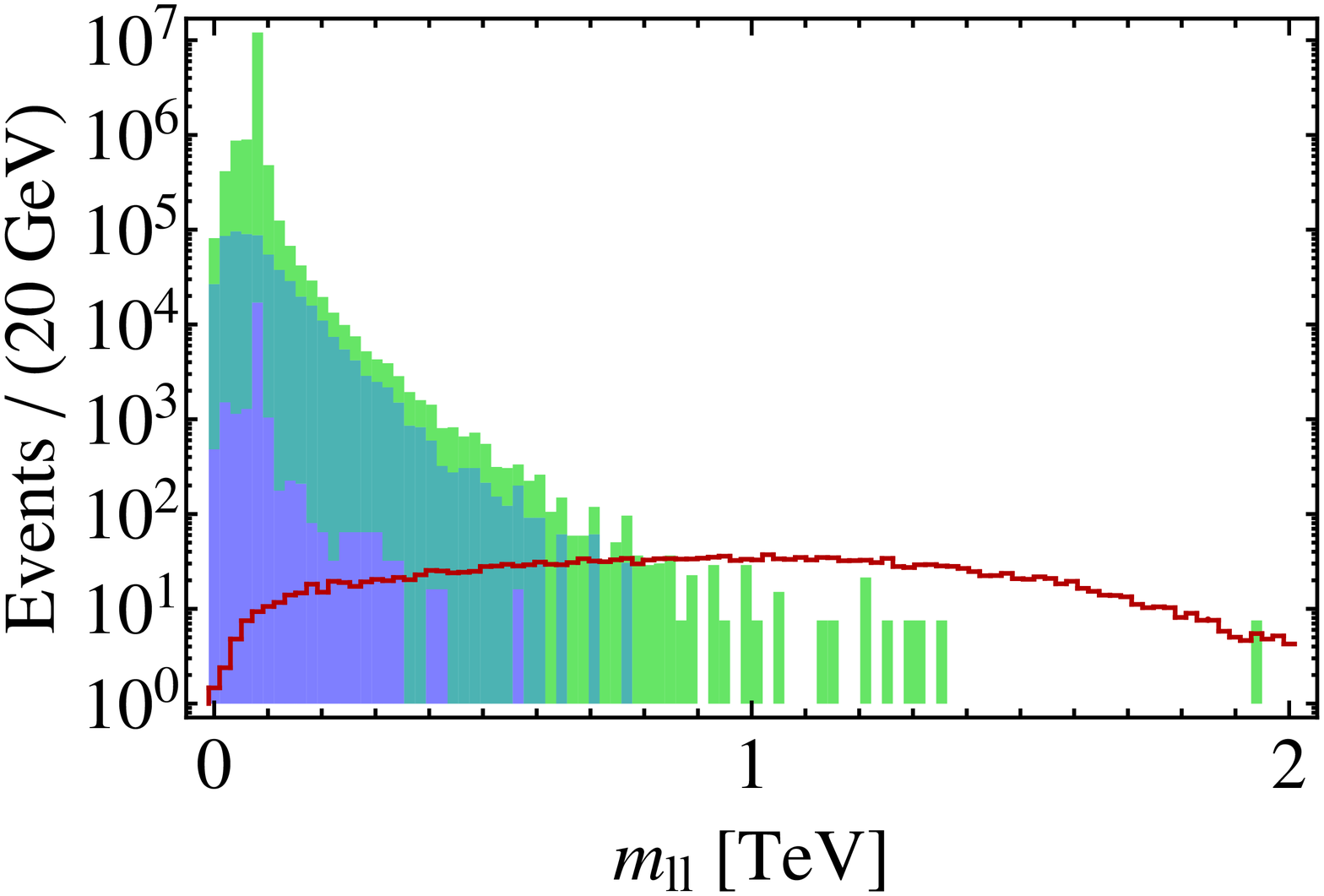}
}
\subfigure[]{
\includegraphics[clip,width=0.458\textwidth]{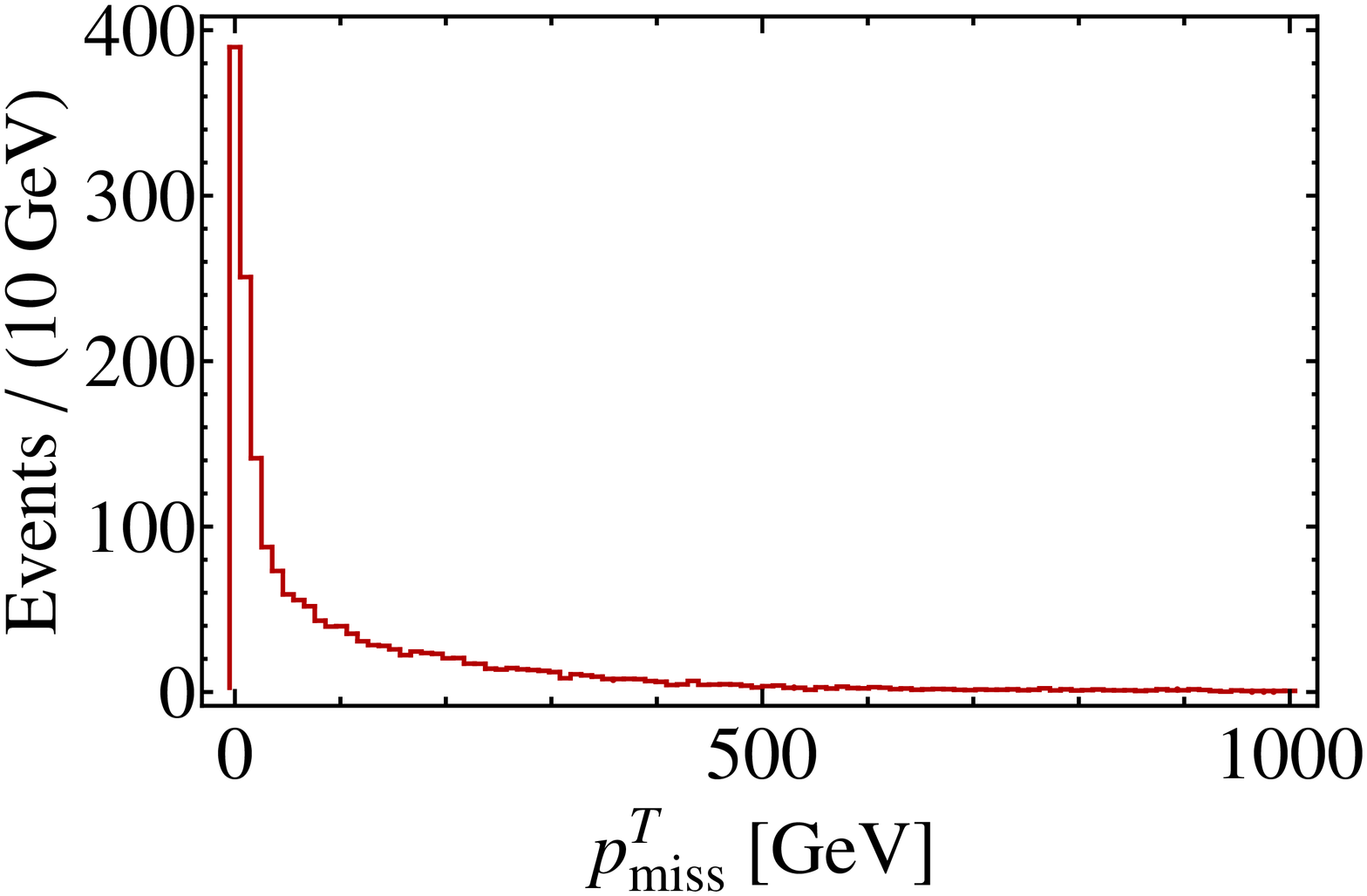}
}
\caption{Distribution of signal events (red) in the invariant masses of the particle combinations $jj \ell_1 \ell_2$ (a), $jj\ell_2$ (b), $\ell_1 \ell_2$ (c), as well as the missing transverse momentum (d) in the benchmark scenario Eq.~(\ref{eq:scenario}), at the LHC with $14$~TeV and ${\mathcal L}=30~{\rm fb}^{-1}.$ We denote with $\ell_{1(2)}$ the leptons with the highest (second highest) $p_T$. In (c), stacked background events are shown for $Z+$~jets (light green), $t\bar t$ (dark green), and $W + Z$ (blue) production.
}
\label{fig:distromass} 
\end{figure}
\begin{table}[t]
\renewcommand{\arraystretch}{1.3}
\vspace{1cm}
\begin{center}
\begin{tabular}{cc}
	\hline
	number of jets              & $N_j             \ge 2$     \\
	number of isolated leptons  & $N_\ell          =   2$     \\
	invariant dilepton mass     & $m_{\ell\ell}    > 0.3$~TeV \\
	total invariant mass        & $m_{\ell\ell jj} > 1.5$~TeV \\
	\hline
\end{tabular}
\end{center}
\renewcommand{\arraystretch}{1.0}
\caption{Selection cuts used in the LHC event analysis.}
\label{tab:selectioncuts}
\end{table}

\subsubsection{Detector Simulation, Event Reconstruction and Selection Cuts}
\label{sec:reco}

To simulate a generic LHC detector, we use the fast detector simulation package {\tt AcerDET}-1.0~\cite{RichterWas:2002ch} with standard settings, for the simulated signal and background events. As we are interested in lepton flavour violating signals, we generate the event sample using a maximal mixing of the active right-handed neutrino to both electrons and muons, but without coupling to taus, i.e. we have $V_{Ne} = V_{N\mu} = 1/\sqrt{2}$ and $V_{N\tau} = 0$.
For the primary selection, we require two {\tt AcerDET} reconstructed isolated leptons ($\ell=e, \mu$) and at least two jets. We show the invariant mass distributions of signal and background events before applying cuts in Figure~\ref{fig:distromass}.

We then apply our selection cuts as summarized in Table~\ref{tab:selectioncuts}, to reduce the background~\cite{Ferrari:2000sp, Gninenko:2006br}. Dilepton pairs from $Z$+jets events are efficiently reduced by requiring a large dilepton mass $m_{\ell\ell}>300$~GeV. A cut well above the $Z$-peak is
necessary, due to the long tail in the $Z\to\ell\ell$ invariant mass distribution, see Figure~\ref{fig:distromass}. In addition, a generous cut on the total invariant mass $m_{\ell\ell jj}>1.5$~TeV reduces $t\bar t$ and $Z$+jets further, without reducing the signal too much, which peaks at $m_{\ell\ell jj}\approx m_{W_R}$, typically with a width of order $100$~GeV due to final gluon radiation and smearing. In Figure~\ref{fig:distromass}, we also show the signal distribution with respect to the missing transverse momentum, but only for illustration, as a cut on this variable does not provide a noticeable improvement of the signal over background ratio. Nevertheless, appropriate selection criteria with respect to the missing momentum can be useful, e.g. in order to decrease possible containment of LNV and LFV signals with SM background processes containing light neutrinos.

\begin{table}[t]
	\renewcommand{\arraystretch}{1.3}
	\centering
	\begin{tabular}{r|rr|rr|rr|rrrr}
	& \multicolumn{4}{|c|}{OS} 
	& \multicolumn{6}{|c }{SS} \\ 
	& $e^+e^-$     & $\mu^+\mu^-$ & $e^+\mu^-$ & $e^-  \mu^+$ &
     $e^+\mu^+$   & $e^-  \mu^-$ & $e^+e^+$   & $e^-e^-$     &
     $\mu^+\mu^+$ & $\mu^-\mu^-$                             \\ 
	\hline
	$t\bar t$ & 190 & 170 & 149 & 164
	& $\approx\!10$ & $\approx\!10$ & $\approx\!10$ & $\approx\!10$ & $\approx\!10$ & $\approx\!10$ \\
	$Zj(j)$     & 181 & 187 & 0 & 2 & 0 & 0 & $\approx\!10$ & $\approx\!10$ & $\approx\!10$ & $\approx\!10$ \\
	\hline
	Signal     & 289 & 192 & 228 & 230 & 330 & 108 & 204 & 74 & 146 & 45 \\
	Eff. [\%]  &  51 & 33 & 42 & 43 & 41 & 41 & 49 & 50 & 35 & 32
\end{tabular}
\renewcommand{\arraystretch}{1.0}
\caption{Background and signal events after cuts, Table~\ref{tab:selectioncuts}, 
for our benchmark scenario, Eq.~(\ref{eq:scenario}), with maximal mixing to electrons and muons, $V_{Ne} = V_{N\mu} = 1/\sqrt{2}, V_{N\tau} = 0$, at the LHC with 
$14$~TeV and ${\mathcal L}=30~{\rm fb}^{-1}$. The dileptons are grouped 
into pairs with same-sign (SS) and opposite-sign (OS) charges. The background event rates for the same-sign signatures were estimated using a 5\% probability of mis-identifying the charge of one of the leptons. For the signal, we give the efficiencies (Eff.) as ratios of events after and before reconstruction and selection cuts in percent.}
\label{tab:CutFlow}
\end{table}
For our benchmark scenario, we summarize the background and signal
events after cuts in Table~\ref{tab:CutFlow}. With these cuts, we
obtain signal efficiencies of $\epsilon \approx 50\%,43\%, 34\%$ for the
dilepton flavour compositions $\ell\ell^\prime = ee,e\mu,\mu\mu $,
respectively. Note that typically the {\tt AcerDET} reconstruction
efficiency is higher for electrons than muons. In addition, the {\tt AcerDET} algorithms are not including inefficiencies for the $e$ and $\mu$ reconstruction. Additional weighting factors of $70$ - $90\%$ for each
lepton could be applied. On the other hand, we do not include a factor
$k\approx1.3$ for the signal events~\cite{Nadolsky:2004vt}. In
Table~\ref{tab:CutFlow}, we have also grouped the dileptons into pairs
with same-sign~(SS) and opposite-sign~(OS) charges. The background rates of the same-sign signatures were estimated using a 5\% probability of mis-identifying the charge of one of the leptons in the respective opposite-sign signature~\cite{Ferrari:2000sp}. No other sources for SS background were taken into account. For $W_R$ masses of the order of TeV, the valence quarks play a significant role for its production, and thus, reflected by the PDFs, the rate for $W_R^+$ production is larger than that for $W_R^-$~\cite{Ferrari:2000sp}. Typically the fraction $W_R^+/(W_R^+ + W_R^-)$ of produced $W_R^+$ in $pp$ collisions at $14$~TeV changes from $\approx 70\%$ to $\approx 95\%$ for $m_{W_R}$ increasing from $1$~TeV to $10$~TeV~\cite{Ferrari:2000sp, Gninenko:2006br}.  We thus obtain for the reconstructed signal events $N(e^+e^-) : N(e^+e^+) : N(e^-e^-) \approx N(\mu^+\mu^-) : N(\mu^+\mu^+):N(\mu^-\mu^-) \approx N(e^+\mu^- + e^-\mu^+) : N(e^+\mu^+) : N(e^-\mu^-) \approx 4 : 3 : 1$, in our benchmark scenario.

\subsubsection{Selection Cut Optimization}

For the purpose of our study, we do not optimize the cuts for each LRSM parameter point defined by $m_{W_R}$ and $m_{N_R}$. Instead, we always use the cuts as given in Table~\ref{tab:selectioncuts}, which were chosen to highlight the discovery reach for the LHC. The cuts could be adapted by requiring higher transverse momenta of the leptons and jets, $p_{\rm T}\gsim 100$~GeV as well as selecting on missing $p_{\rm T}^{\rm miss}<50$~GeV~\cite{delAguila:2009bb}. Further, one can apply mass-window cuts to isolate the $W_R$ and $N_R$ mass peaks in the $m_{\ell jj}$ and $m_{\ell\ell jj}$ invariant mass distributions, respectively~\cite{Ferrari:2000sp, Gninenko:2006br}. In particular, different $m_{\ell jj}$ mass-window cuts might be necessary to disentangle different signal contributions, if more than one $N_R$ is kinematically accessible. Also, our cut on the total invariant mass $m_{\ell \ell j j} > 1.5$~TeV is chosen for heavy bosons $m_{W_R} \gsim 2$~TeV, and for smaller masses one loses signal events, which can be observed in the indentation on the left side of the $10^3$ events contour in Figure~\ref{fig:MW_MN_ee}.

The cuts might also be improved in the kinematically suppressed parameter regions, i.e. for $m_{N_R} \lsim m_{W_R}$ (threshold region) and $m_{N_R} \ll m_{W_R}$ (jet region). To enhance the sensitivity in the threshold region, same-sign lepton pairs can be
selected, or the cuts can be raised for the $p_T$ of the jets and for the dilepton invariant mass $m_{\ell\ell}$~\cite{Ferrari:2000sp}. In the jet region, where the lepton from the $N_R$ decay tends to lie inside the jet cones, one could search for events with one high $p_T$ isolated lepton and one high $p_T$ hadronic jet with a large electromagnetic component, and matching a high $p_T$
track in the inner detector. This typically increases the sensitivity in those regions by a few percent~\cite{Ferrari:2000sp, Gninenko:2006br}.

For same-sign dileptons, the kinematic cuts can be relaxed, since SM background events will dominantly produce opposite-sign dileptons. As mentioned above, same-sign lepton background events would originate from charge mis-identification, mistakenly reconstructed leptons from jets and SM diboson production~\cite{Aad:2011vj}. The dilepton pairs could also be grouped according to their flavour configurations; for example, lepton pairs with different flavours only originate from $t\bar t$ production, reducible by vetoing $b$-jets. Since the b-tagging efficiencies at the LHC are of the order of $60$\% to $70$\%, the $t\bar t$ background should be reduced by an order of magnitude. The signal loss will be small, since we typically have $Br(N_R \to \ell b t)\approx 20\%$, and $Br(t\to b\ell\nu) = 20\%$, $\ell = e,\mu$. However, as pointed out earlier, to estimate the discovery potential of the LHC for lepton flavour violating signals, we will base the following analysis on the basic conservative kinematic cuts as given in Table~\ref{tab:selectioncuts}.

\subsection{Lepton Flavour Violation}\label{sec:LeptonFlavourViolation}

In this section we discuss the prospects of observing lepton flavour
violating production and decay of right-handed $W_R$ bosons and heavy
neutrinos in the left-right symmetric model described in
Section~\ref{sec:minimalLR}, due to a mixing among the right-handed
neutrinos. We will focus on a mixing of right-handed neutrinos with electrons and muons, i.e. we are mostly interested in the mixing matrix elements $V_{N_i e}$ and $V_{N_i \mu}$. In turn, we will assume the case of flavour mixing with the unitarity constraints $V_{N_i e}^2 + V_{N_i \mu}^2 = 1$ ($i=1,2$), $V_{N_1 e}V_{N_2 e} + V_{N_1 \mu}V_{N_2 \mu} = 0$, with either one or two heavy neutrinos light enough to be produced at the LHC\footnote{We always assume real mixing matrix elements, i.e. we neglect any possible CP violating phases in the right-handed neutrino sector.}. In the first case, we will also discuss the more general scenario of a possible mixing to taus by extending the unitarity relation as $V_{N_1 e}^2 + V_{N_1 \mu}^2 + V_{N_1 \tau}^2 = 1$. Here, we will not take into account taus in the final state\footnote{This is a highly interesting possibility in its own right as the reconstruction efficiencies in the leptonic channels $\tau \to \ell \bar\nu_\ell \nu_\tau$ with $\ell=e, \mu$ are only reduced by about a third of those for electrons and muons, simply due to the leptonic branching ratios of the $\tau$~\cite{AguilarSaavedra:2012fu}. This could make it possible to detect $\tau-\mu$ and $\tau-e$ flavour violation at the LHC.}, but a non-zero value of $V_{N_1 \tau}$ will reduce the mixing to electrons and muons accordingly. 

\subsubsection{Single Neutrino Exchange}
\label{sec:SingleNeutrino}

To simplify our discussion we first consider one right-handed neutrino
in the intermediate state. That is, either only one right-handed neutrino
is light enough to be produced, $m_{N_1} < m_{W_R} < m_{N_{2,3}}$, or if more than one neutrino is below the threshold, the mass difference between the right-handed neutrinos is sufficiently large, $\Delta m_{N_i N_j} \gsim 100$~GeV, such that the neutrino resonances can be individually reconstructed. In Section~\ref{sec:TwoNeutrinos}, we then discuss right-handed neutrinos with smaller mass differences.

\begin{figure}[t]
\centering
\includegraphics[clip,width=0.47\textwidth]{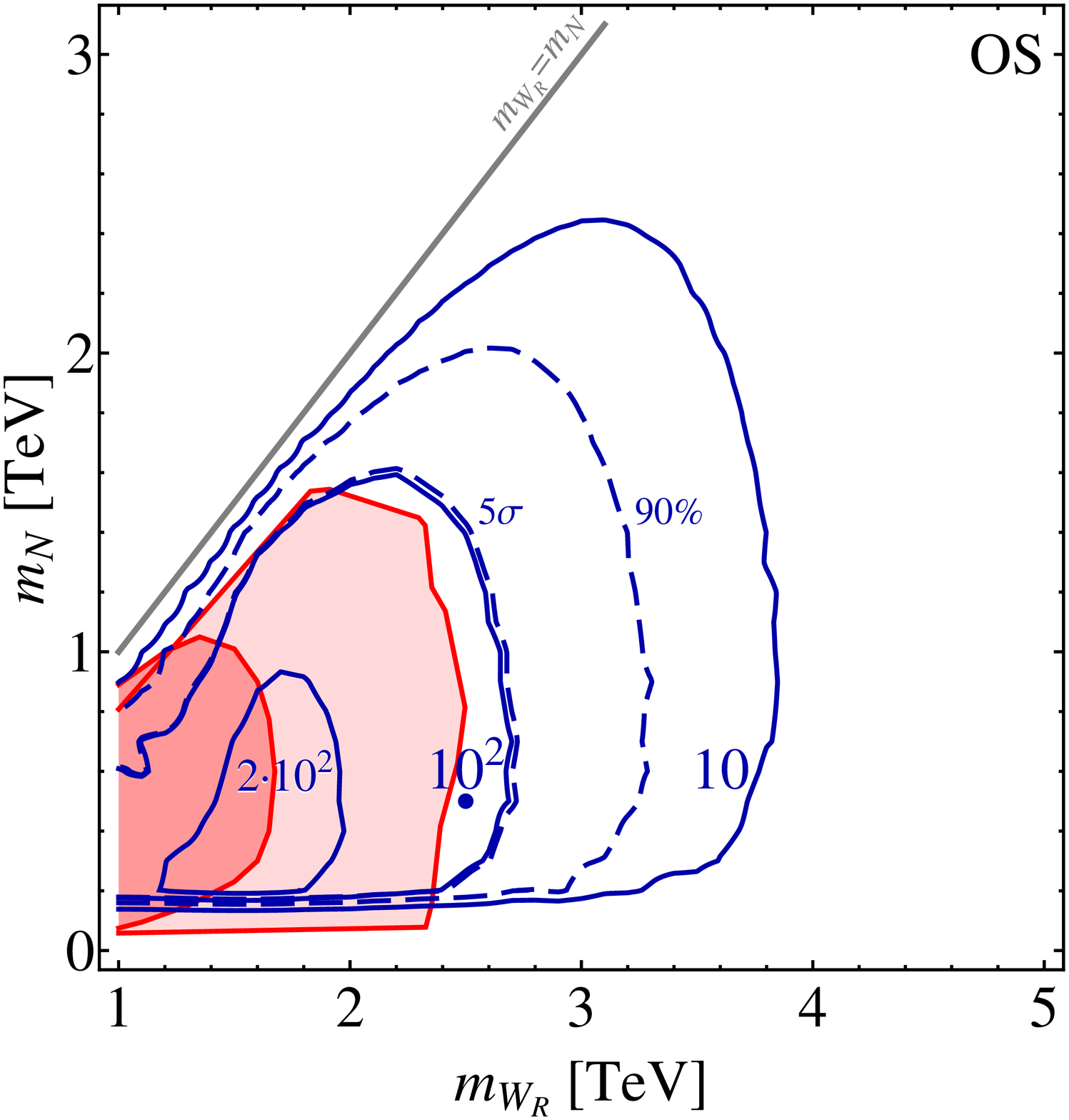}
\includegraphics[clip,width=0.47\textwidth]{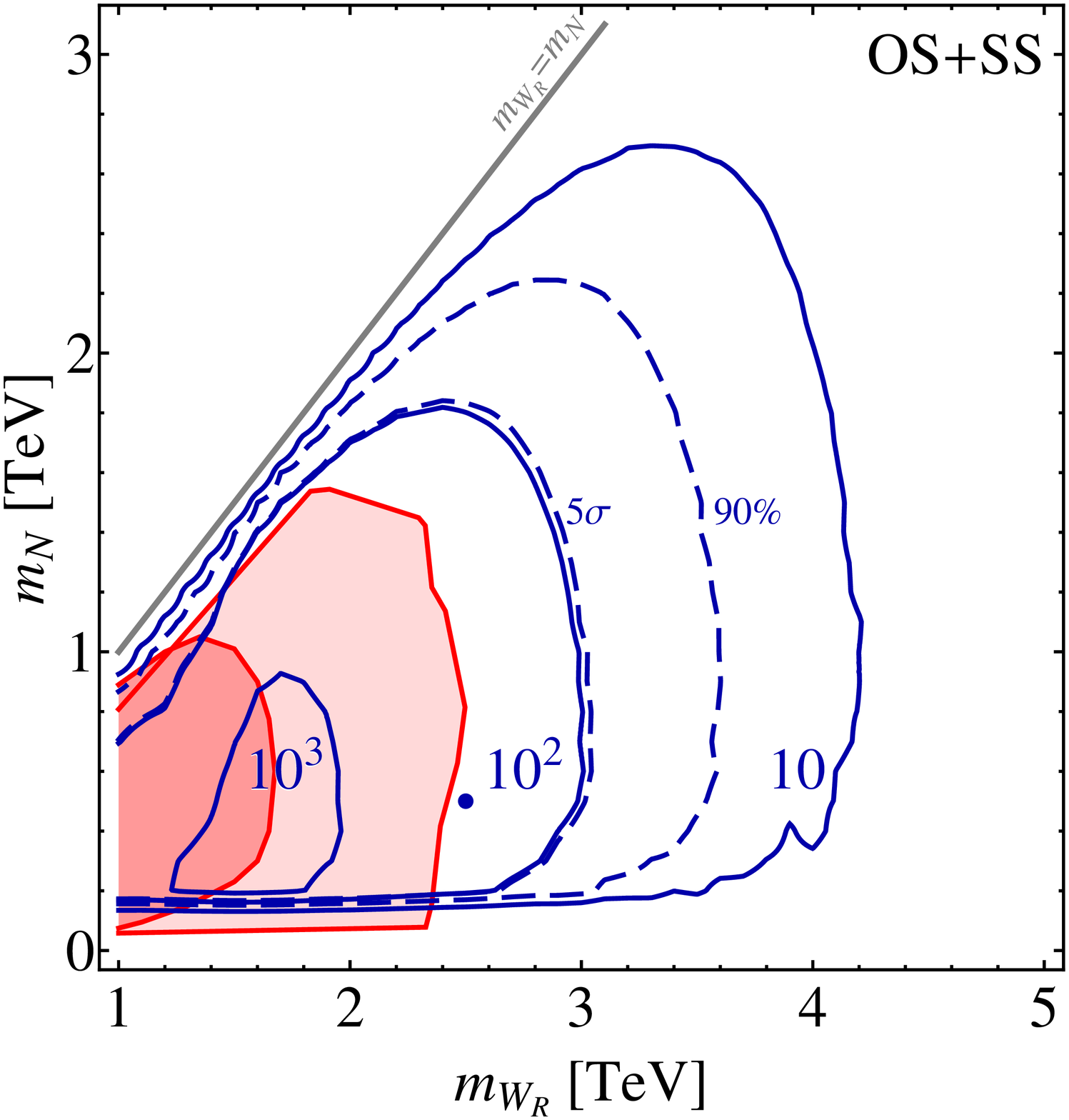}
\caption{The solid blue contours denote the event rates for the processes $pp\to W_R
  \to e^\pm \mu^{\mp} + 2\text{ jets}$ (left) and $pp\to W_R \to e^\pm
  \mu^{\pm,\mp} + 2\text{ jets}$ (right), as a function of
  right-handed $W$ boson mass $m_{W_R}$, and the right-handed neutrino
  mass $m_N$, at the LHC with 14~TeV and $\mathcal{L} = 30\text{
    fb}^{-1}$. The grey contour corresponds to the kinematical
  threshold $m_{W_R} = m_N$. The blue dashed contours
  give the discovery and exclusion reach with $S/\sqrt{B}=5$ and
  $1.64$ (90\% C.L.), respectively. The processes are calculated for
  maximal unitary coupling of the right-handed neutrino to $e$ and
  $\mu$ only with $V_{Ne} = V_{N\mu} = 1/\sqrt{2}$. The red shaded areas are excluded by current LHC searches at CMS~\cite{CMS:925203} (dark shaded) and ATLAS~\cite{CERN-PH-EP-2012-022}~(light shaded).}
\label{fig:MW_MN_emumax} 
\end{figure}
The couplings of the right-handed neutrino $N_1 \equiv N$ to electrons
and muons are given by the elements $V_{Ne}$ and $V_{N\mu}$ of the
right-handed neutrino mixing matrix. After reconstruction and cuts,
the number of opposite-sign~(OS) dilepton events from the processes
$pp\to W_R \to e^+e^- (\mu^+\mu^-, e^\pm\mu^\mp) + 2\text{ jets}$, are
given by
\begin{align}
\label{eq:SimplifiedLFVEventRates}
	N(e^+  e^-)     &= \epsilon^\text{\tiny OS}_{ee} \times \mathcal{L} \times
		\sigma_{ee}^\text{\tiny OS}\left( m_{W_R}, m_N \right) 
		\times |V_{Ne}|^4,    \nonumber\\
	N(\mu^+\mu^-)   &= \epsilon^\text{\tiny OS}_{\mu\mu} \times\mathcal{L} \times
		\sigma_{ee}^\text{\tiny OS}\left( m_{W_R}, m_N \right) 
		\times |V_{N\mu}|^4,          \\
	N(e^\pm\mu^\mp) &= \epsilon^\text{\tiny OS}_{e\mu} \times \mathcal{L} \times 
		\sigma_{ee}^\text{\tiny OS}\left( m_{W_R}, m_N \right)
		\times |V_{Ne}|^2 |V_{N\mu}|^2, \nonumber
\end{align}
with the integrated luminosity $\mathcal{L}$, the cross section
$\sigma_{ee}^\text{\tiny OS}\left( m_{W_R}, m_N \right)$ for $e^+e^-$ production with $V_{Ne}=1$ (no flavour mixing), and the experimental efficiencies $\epsilon^\text{\tiny OS}_{ee}$, $\epsilon^\text{\tiny OS}_{\mu\mu}$, $\epsilon^\text{\tiny OS}_{e\mu}$ due to reconstruction and cuts, for the different dilepton flavour combinations. Similar relations are obtained for same-sign dilepton pairs, with the efficiencies $\epsilon^\text{\tiny SS}_{ee}$, $\epsilon^\text{\tiny SS}_{\mu\mu}$, $\epsilon^\text{\tiny SS}_{e\mu}$ and the SS cross section $\sigma_{ee}^\text{\tiny SS}\left( m_{W_R}, m_N \right)$.  The numerical values of the efficiencies are summarized in Table~\ref{tab:CutFlow}, obtained in our simulation of the benchmark scenario $m_{W_R} =
2$~TeV, $m_{N} = 0.5$~TeV with the cuts shown in Table~\ref{tab:selectioncuts}.

To determine the reach of the LHC to probe lepton flavour violation,
we show in Figure~\ref{fig:MW_MN_emumax} the event rate for the
signature $pp\to W_R \to e\mu + 2\text{ jets}$ at the LHC with 14~TeV
and an integrated luminosity of $\mathcal{L} = 30\text{ fb}^{-1}$ for
maximal mixing $V_{Ne} = V_{N\mu} = 1/\sqrt{2}$. In
Figure~\ref{fig:MW_MN_emumax} (left) we show the sum of events of
opposite-sign leptons, $N(e^+\mu^-) + N(e^-\mu^+)$, whereas in the
right plot the charges are summed over as well, $N(e^+\mu^-) + N(e^-\mu^+) +
N(e^+\mu^+) + N(e^-\mu^-)$. The blue dashed contours show the
discovery reach ($S/\sqrt{B}=5$, practically coinciding with the 100
events contour) and the exclusion reach ($90\%$~C.L. excess over
background). As discussed previously, we do not optimize the cuts for
each individual parameter point, which could extend the reach
somewhat. Because of the simple relations in
Eq.~(\ref{eq:SimplifiedLFVEventRates}), the LHC reach is comparable
for lepton flavour conserving and maximally violating signals,
c.f. Figures~\ref{fig:MW_MN_ee} and
~\ref{fig:MW_MN_emumax}. Effectively, the rate is halved for maximal
mixing, and the shape of the regions follows from the discussion in
Section~\ref{sec:ResultsDilepton}. 
It should
therefore be possible to probe LFV for $W_R$ masses up to 3-3.5~TeV
and right-handed neutrino masses up to 1.5-2~TeV. For comparison, we show the CMS exclusion region with $\mathcal{L}=0.24\text{ fb}^{-1}$~\cite{CMS:925203}, and the ATLAS exclusion region with $\mathcal{L}=2.1\text{ fb}^{-1}$~\cite{CERN-PH-EP-2012-022}.
\begin{figure}[t]
\centering
\includegraphics[clip,width=0.49\textwidth]{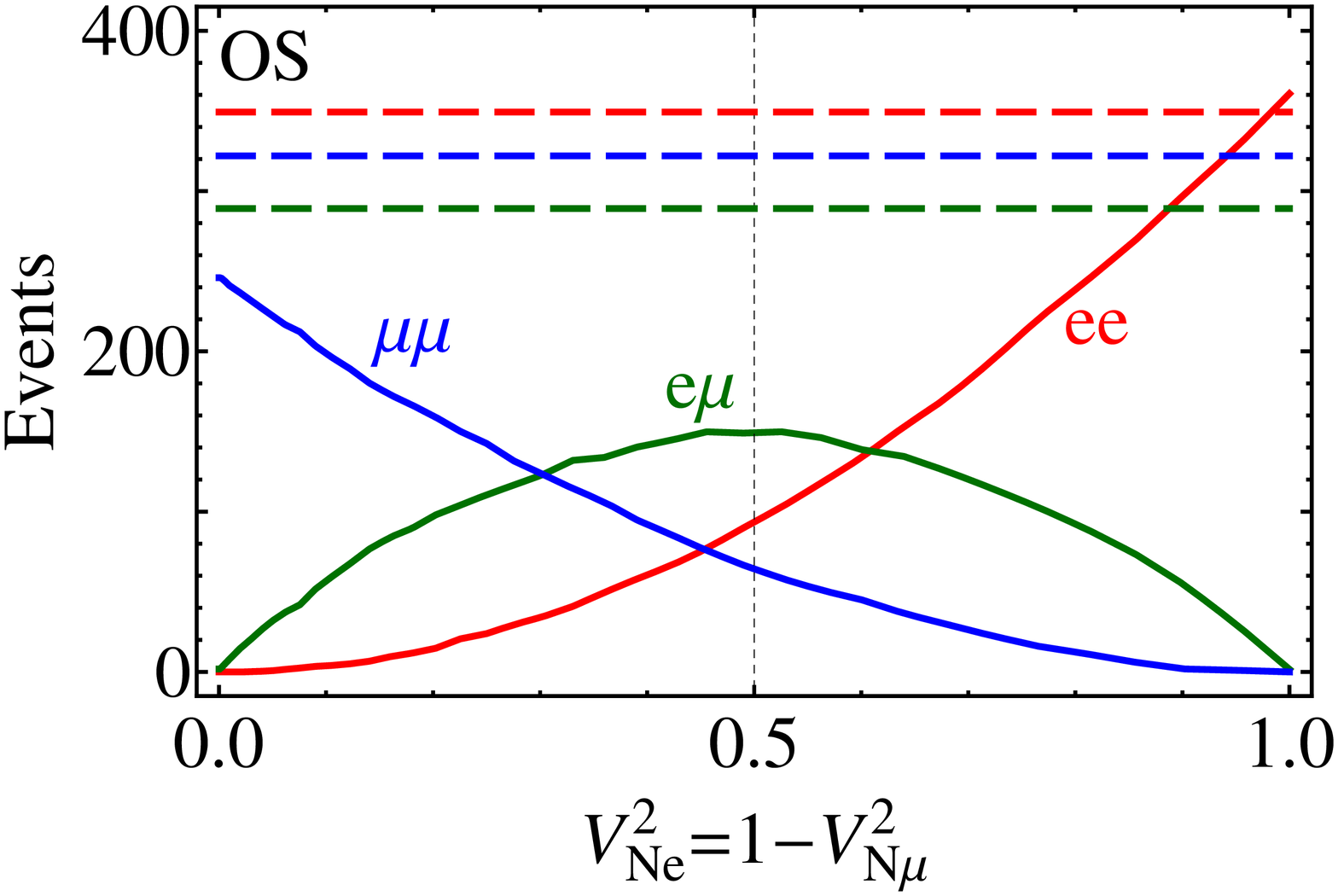}
\includegraphics[clip,width=0.49\textwidth]{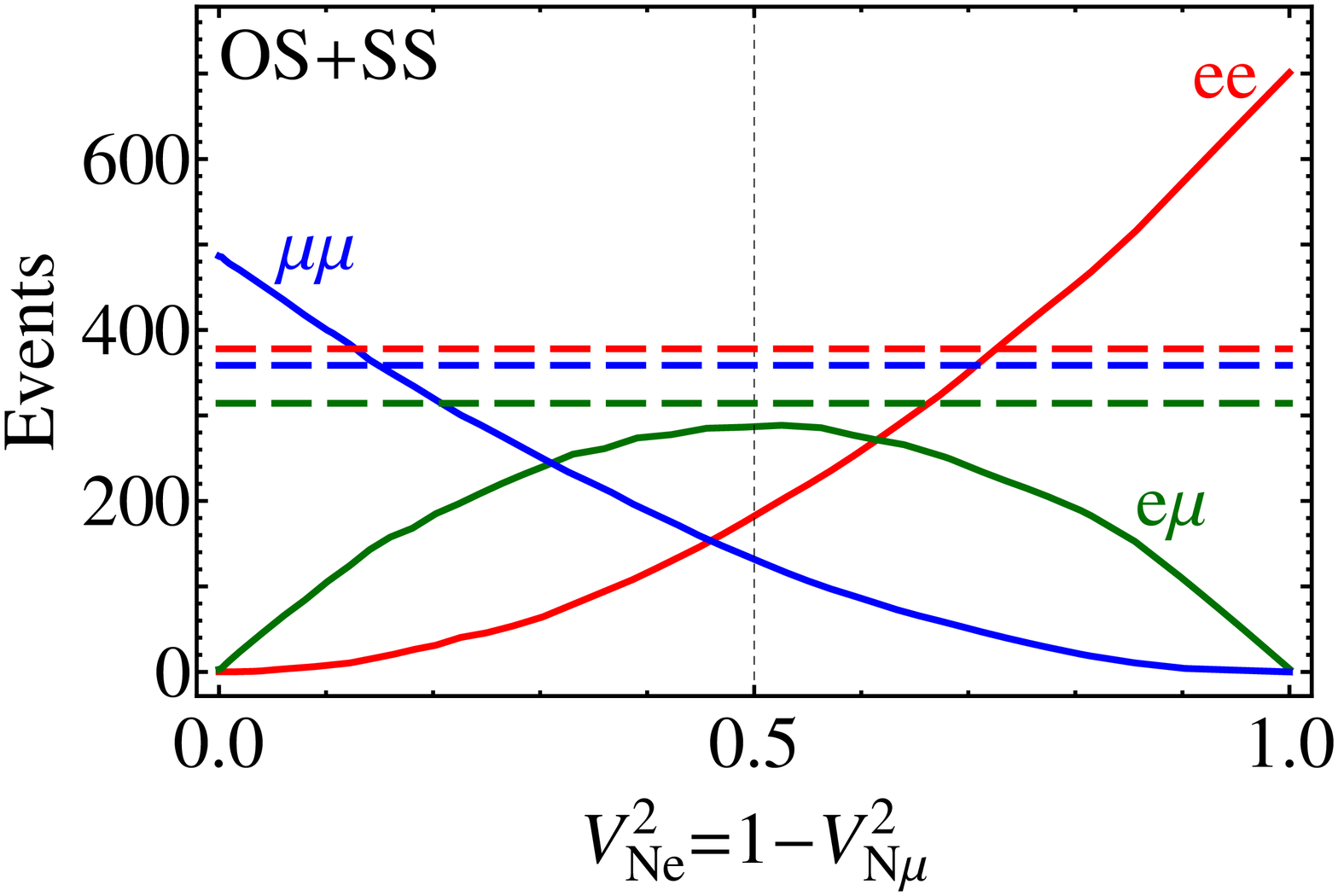}
\caption{Event rates for the processes $pp\to W_R \to e e (\mu\mu,
  e\mu) + 2\text{ jets}$ at the LHC with 14~TeV and $\mathcal{L} =
  30\text{ fb}^{-1}$ as a function of the right-handed neutrino-electron
  coupling $V^2_{Ne} = 1 - V_{N\mu}^2$ for a right-handed $W$
  boson mass $m_{W_R}=2.5$~TeV and right-handed neutrino mass
  $m_N=0.5$~TeV. The rates are calculated using only the opposite
  charge sign event sample (left) and both opposite and same charge
  sign sample (right) of the two leptons. The horizontal dashed lines show the background to each correspondingly colored signal.}
\label{fig:Veos} 
\end{figure}

To verify the LHC sensitivity to the flavour couplings $V_{Ne}$ and
$V_{N\mu}$, we fix the masses $m_{W_R}=2.5$~TeV, $m_N=0.5$~TeV, and show
in Figure~\ref{fig:Veos} the event rates as a function of the electron
coupling $V_{Ne}$ for OS and OS+SS signatures, where we assume
unitarity among the couplings $V_{Ne}$ and $V_{N\mu}$, i.e. $V^2_{Ne}
+ V_{N\mu}^2 = 1$. The dependence of the event rates on the couplings
is again very simple and follows
Eqs.~(\ref{eq:SimplifiedLFVEventRates}). If the lepton flavour
violating signature $e$-$\mu$ is observable, at least one of the
flavour conserving signatures $e$-$e$ or $\mu$-$\mu$ can be observed
as well. If $V_{Ne}^2 > 1/2$, the neutrino couples more strongly to
electrons than to muons, and from
Eqs.~(\ref{eq:SimplifiedLFVEventRates}) we can determine
\begin{equation}
\label{eq:VNeFromRates}
	V_{Ne}^2 = 
		\left[
			1+\frac{1}{2}\frac{\epsilon^{OS}_{ee}}{\epsilon^{OS}_{e\mu}}
			\frac{N(e^+\mu^- + e^-\mu^+)}{N(e^+e^-)}
		\right]^{-1},
\end{equation}
for opposite-sign signatures. The coupling parameter $|V_{Ne}|$ can
therefore be determined independently of the absolute normalization of
the production cross section (e.g. due to possible splitting of the gauge couplings, $g_R \neq g_L$
effects) and systematic uncertainties are reduced. If $V_{Ne}^2 <
1/2$, we find an analogous expression, replacing $N(e^+e^-)$ with
$N(\mu^+\mu^-)$ in the denominator of Eq.~(\ref{eq:VNeFromRates}). For
$V_{Ne}^2 \approx 1/2$, Eq.~(\ref{eq:VNeFromRates}) can be used both
with $N(e^+e^-)$ and $N(\mu^+\mu^-)$, yielding two independent
measurements for $V_{Ne}^2$.

\begin{figure}[t]
\centering
\includegraphics[clip,width=0.50\textwidth]{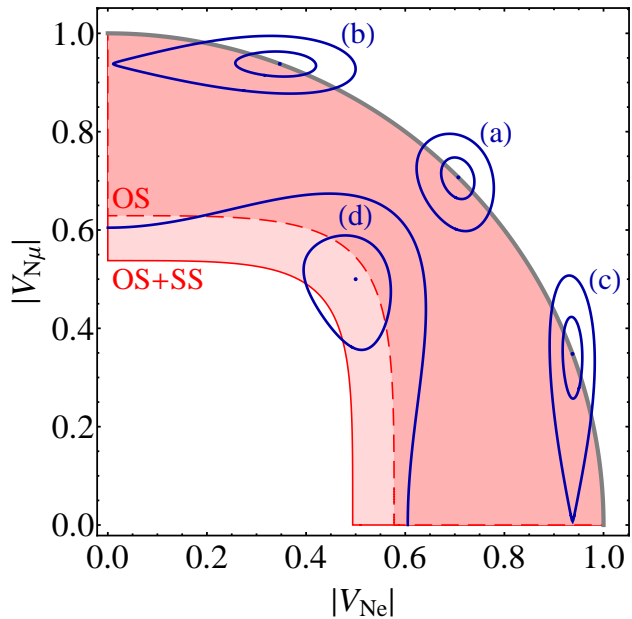}
\caption{Excluded regions in the $V_{Ne}$ and $V_{N\mu}$ parameter plane at 90\% C.L. using OS (light red shaded with solid boundary) and OS+SS (dark red with dashed boundary) event samples, assuming no excess above background. Also shown are the $1\sigma$ and $5\sigma$ uncertainty solid blue contours for hypothetical signals corresponding to $(V_{Ne},V_{N\mu}) = (1/\sqrt{2}, 1/\sqrt{2})$, $(0.347, 0.938)$, $(0.937, 0.348)$, $(0.5, 0.5)$ using OS+SS samples. The statistical analysis is based on the events $pp \to W_R \to e e (\mu \mu, e \mu) + 2\text{ jets}$ at the LHC with 14~TeV and $\mathcal{L} = 30\text{ fb}^{-1}$, with a $W_R$ boson mass of $m_{W_R}=2.5$~TeV and right-handed neutrino mass of $m_N=0.5$~TeV. The errors on the event rates are assumed to be dominated by their statistical uncertainty.}
\label{fig:VeVmu} 
\end{figure}
This can be generalized by dropping the unitarity condition and
keeping the couplings $V_{Ne}$ and $V_{N\mu}$ independent from each other. This
approach is used in Figure~\ref{fig:VeVmu}, where we show the excluded
region in the $V_{Ne}$ and $V_{N\mu}$ parameter plane at 90\% C.L.,
using a $\chi^2$ analysis of the $ee$, $e\mu$ and $\mu\mu$ events with
both OS and OS+SS signatures, assuming no excess above the background is observed. In the given scenario with
$m_{W_R}=2.5$~TeV and $m_N=0.5$~TeV, couplings of the order of $V_{Ne(\mu)} \approx 0.5$ can be
excluded. For illustration, we also show uncertainty contours for
hypothetical signals corresponding to four different choices for
$(V_{Ne}, V_{N\mu})$: (a) $(1/\sqrt{2},1/\sqrt{2})$, corresponding to
maximal unitary mixing; (b) $(0.347,0.938)$, representing
the unitary scenario with the minimal value of $V_{Ne}$ that can be
distinguished from zero; (c) $(0.937,0.348)$, representing the unitary
scenario with the minimal value of $V_{N\mu}$ that can be distinguished
from zero; (d) $(0.5,0.5)$, representing a non-unitary
scenario close to the exclusion limit. In Figure~\ref{fig:VeVmu}, the errors on the event rates are
assumed to be dominated by their statistical uncertainties.

\begin{figure}[t]
\centering
\includegraphics[clip,width=0.49\textwidth]{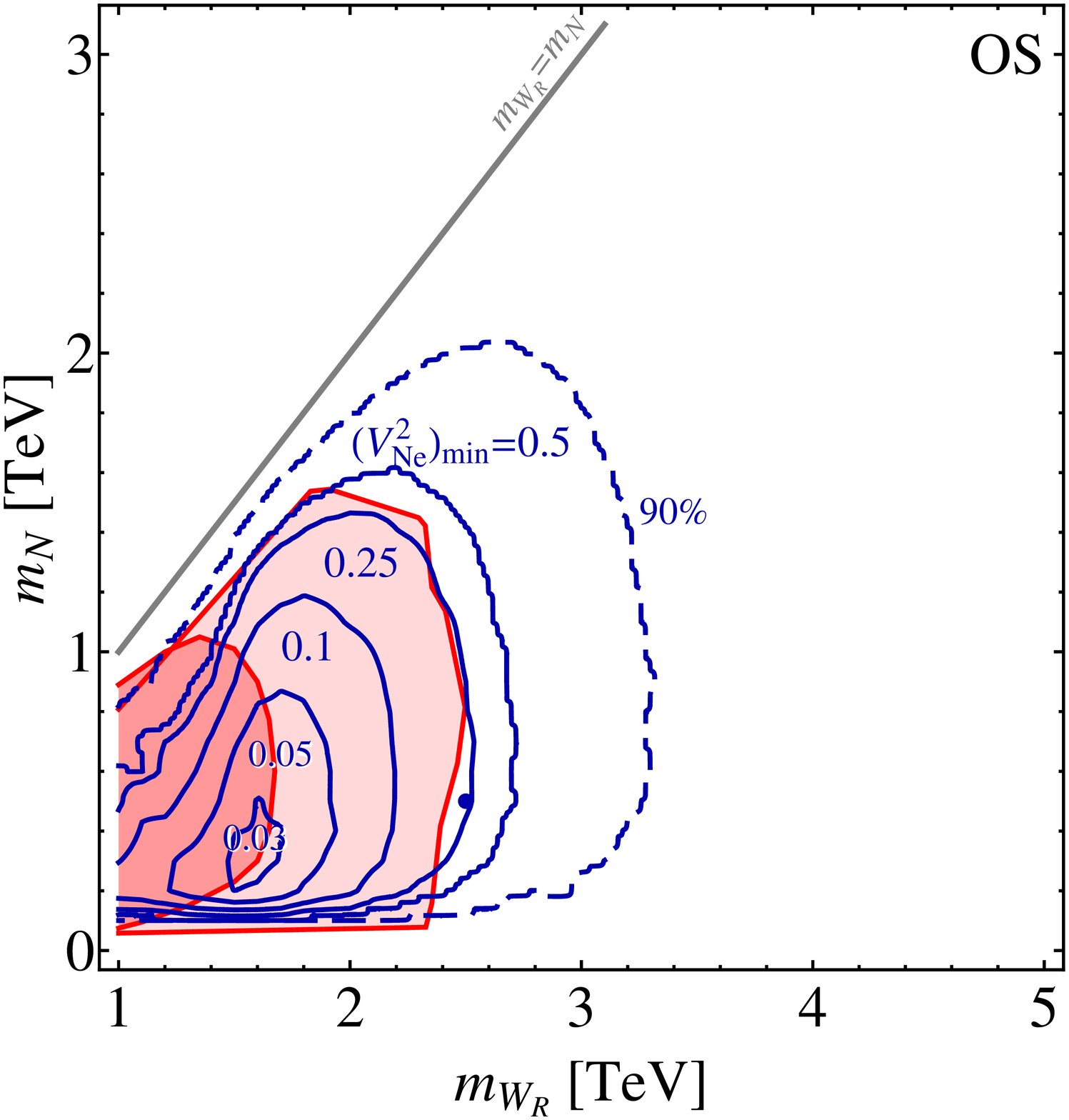}
\includegraphics[clip,width=0.49\textwidth]{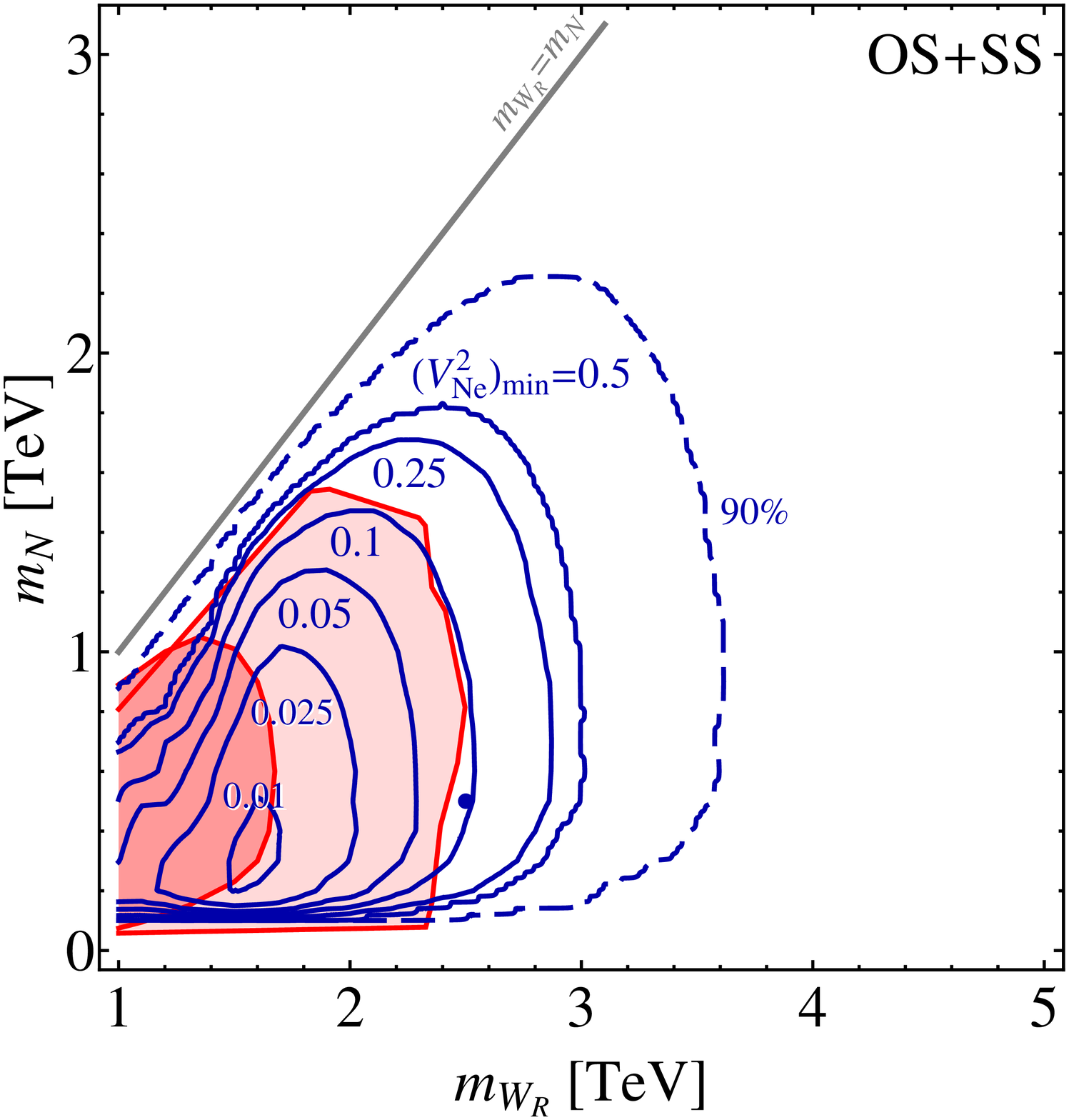}
\caption{Sensitivity to the coupling $V^2_{Ne}$ as function of $m_{W_R}$ and $m_{N_R}$ at the LHC with 14~TeV and $\mathcal{L}=30\text{ fb}^{-1}$ using OS (left) and OS+SS leptons (right). The solid contours indicate a discovery sensitivity at $5\sigma$ and the outermost contour corresponds to an exclusion at 90\% C.L. for maximal mixing $V_{Ne}^2 = 1/2$. The grey contour corresponds to the kinematical
  threshold $m_{W_R} = m_N$. The red shaded areas are excluded by current LHC searches at CMS~\cite{CMS:925203} (dark shaded) and ATLAS~\cite{CERN-PH-EP-2012-022}~(light shaded).}
\label{fig:Ve_MWMN} 
\end{figure}

In order to derive the sensitivity as a function of $m_{W_R}$ and
$m_N$, we show in Figure~\ref{fig:Ve_MWMN} the minimal coupling
$V_{Ne}^2$ which can be observed at $5\sigma$, for both OS and OS+SS
leptons. The outermost contour in the plots corresponds to a signal at
90\% C.L. for maximal mixing $V_{Ne}^2 = V_{N\mu}^2 = 1/2$. This
exclusion contour and the $5\sigma$ discovery contour therefore correspond to
the dashed contours in Figure~\ref{fig:MW_MN_emumax},
respectively. The parameter region with the largest cross section
around $(m_{W_R},m_N)=(1.6,0.3)$~TeV is already excluded by LHC searches, and
flavour violating right-handed neutrino-lepton couplings down to
$V^2_{Ne(\mu)} \approx 10^{-1}$ can potentially be probed at the LHC
with 14~TeV and $\mathcal{L} = 30\text{ fb}^{-1}$.

%

\subsubsection{Two Neutrino Exchange}
\label{sec:TwoNeutrinos}

In general, all heavy neutrinos with $m_{N_i} < m_{W_R}$ that couple
to electrons and/or muons will contribute to our dilepton signatures
and have to be taken into account. When summing over the heavy right-handed
neutrinos in the intermediate state, this may lead to interference
effects between the different contributions. In fact in the limit of
degenerate heavy neutrinos, $\Delta m^2_{ij} \equiv
m^2_{N_i}-m^2_{N_j} \to 0$, all lepton flavour violating signals will
suffer a GIM-like suppression, analogous to low
energy LFV processes as described in Section~\ref{sec:limitsLFV}. As a
crucial difference to the radiative rare decays, the neutrinos are produced on-shell at the LHC with a short decay length. This leads
to a decoherence of the right-handed neutrino oscillation, and the
suppression is proportional to $\Delta m^2_{ij}/(m_N \Gamma_N)$,
rather than $\Delta m^2_{ij}/m_N^2$. This follows from the well
justified narrow width approximation for the product of the neutrino
propagators $N_i$ and $N_j$ in the squared matrix
element~\cite{Deppisch:2003wt}
\begin{eqnarray}
\label{eq:NarrowWidth}
\lefteqn{
	(p^2 - m^2_{N_i} + i m_{N_i} \Gamma_{N_i})^{-1} \times
	(p^2 - m^2_{N_j} - i m_{N_j} \Gamma_{N_j})^{-1} \approx} \nonumber \\ [3mm]&& \qquad \qquad
	\frac{\pi C_{ij}}{2\left< m \Gamma \right>_{ij}}
	\left[
		\delta( p^2 - m^2_{N_i}) + \delta( p^2 - m^2_{N_j})
	\right],
\end{eqnarray}
with
\begin{equation}
	C_{ij} = \left( 1 + i \frac{\Delta m^2_{ij}}{2\left<m\Gamma\right>_{ij}}\right)^{-1},
	\quad
	\Delta m^2_{ij} = m^2_{N_i} - m^2_{N_j},
	\quad
	\left<m\Gamma\right>_{ij} = \frac{1}{2}(m_{N_i}\Gamma_{N_i} + m_{N_j}\Gamma_{N_j}).
\end{equation}
For large mass splittings, $\Delta m^2_{ij} \gg
\left<m\Gamma\right>_{ij}$, the factors $C_{ij}$ approach
$\delta_{ij}$, i.e. the neutrino interference is suppressed and their
contributions add up incoherently in the squared matrix element. For
small mass splittings, $\Delta m^2_{ij} \ll
\left<m\Gamma\right>_{ij}$, due to Eq.~(\ref{eq:NarrowWidth}) and the
unitarity of the mixing matrix $V$ in the vertex
$V_{N_i\ell}N_i$-$\ell$-$W_R$, the flavour violating
process $pp \to W_R \to \ell_1 \ell_2 + 2\text{ jets}$,
$\ell_1 \neq \ell_2$, is suppressed as $\Delta
m^2_{ij}/\left<m\Gamma\right>_{ij}$. When two neutrinos $N_{1,2}$
couple to $e$ and $\mu$, with a unitarity matrix described by the mixing angle $\phi$,
\begin{equation}
\label{eq:twoflavourmix}
	V = 
		\begin{pmatrix}
	 		\cos\phi & \sin\phi \\
			-\sin\phi & \cos\phi
		\end{pmatrix},
\end{equation}
the total event rate for the process $pp \to W_R \to e^\pm \mu^\mp + 2 \text{ jets}$ is then
\begin{equation}
\label{eq:NemuPhiDm}
	N(e^\pm\mu^\mp) \approx \epsilon^\text{\tiny OS}_{e\mu} \mathcal{L} 
		\left[\sigma(m_{W_R},m_{N_1}) + \sigma(m_{W_R},m_{N_2})\right] \times
		\frac{1}{4}\sin^2(2\phi)
		\frac{(\Delta m^2_{12})^2}{(\Delta m^2_{12})^2 + 2\left<m\Gamma\right>^2_{12}}.
\end{equation}
Here, $\sigma\left(m_{W_R},m_{N_i}\right)$ is the cross section of the
lepton flavour conserving process $pp \to e^+ e^- + 2\text{ jets}$ with exchange of one neutrino of mass $m_{N_i}$ coupling only to electrons. This equation describes the event rate for all mass splitting regimes to a very good approximation:
\begin{enumerate}
\item If the neutrino mass difference is larger than the experimental
  width of the neutrino resonances in the invariant mass distribution
  $m_{\ell_2jj}$, $\Delta m_{ij} \gsim \Gamma^{\ell jj}_{N_i}$, it is in principle
  possible to reconstruct each of the resonances $m_{N_{1,2}}$, and to
  determine the couplings $V_{N_1e}^2$, $V_{N_1\mu}^2$,
  $V_{N_2e}^2$, $V_{N_2\mu}^2$ of both neutrinos, as described in
  Section~\ref{sec:SingleNeutrino}. The individual event rate for each
  resonance is then
\[
	N_{N_i}(e^\pm\mu^\mp) \approx \epsilon^\text{\tiny OS}_{e\mu} \mathcal{L} 
		\sigma\left(m_{W_R},m_{N_i}\right) \times \frac{1}{4}\sin^2(2\phi),
\]
which is independent of the neutrino mass splitting.

\item If $\Gamma_{N_i} \ll \Delta m_{ij} \ll \Gamma^{\ell jj}_{N_i}$,
  it is not possible to resolve the individual neutrino contributions
  which overlap and form a single resonance at $m_{\bar N} \approx
  m_{N_1} \approx m_{N_2}$. Nevertheless, the LFV rate is still
  unsuppressed by the mass splitting and the total event rate in the
  single resonance is given by
\[
	N(e^\pm\mu^\mp) \approx \epsilon^\text{\tiny OS}_{e\mu} \mathcal{L} 
		\sigma\left(m_{W_R},m_{\bar N}\right) \times \frac{1}{2}\sin^2(2\phi).
\]

\item If $\Delta m_{ij} \ll \Gamma_{N_i}$, the neutrinos form a single
  resonance at $m_{\bar N}$ and the total event rate is suppressed
  with $\Delta m_{ij}= m_{N_i} - m_{N_j}$ as
\[
	N(e^\pm\mu^\mp) \approx \epsilon^\text{\tiny OS}_{e\mu} \mathcal{L} 
		\sigma\left(m_{W_R},m_{\bar N}\right) \times 
		\sin^2(2\phi)\left(\frac{\Delta m_{ij}}{\langle\Gamma\rangle}\right)^2,
	\quad
	\langle\Gamma\rangle = (\Gamma_{N_i}+\Gamma_{N_j})/2.
\]
\end{enumerate}
\begin{figure}[t!]
\centering
\includegraphics[clip,width=0.62\textwidth]{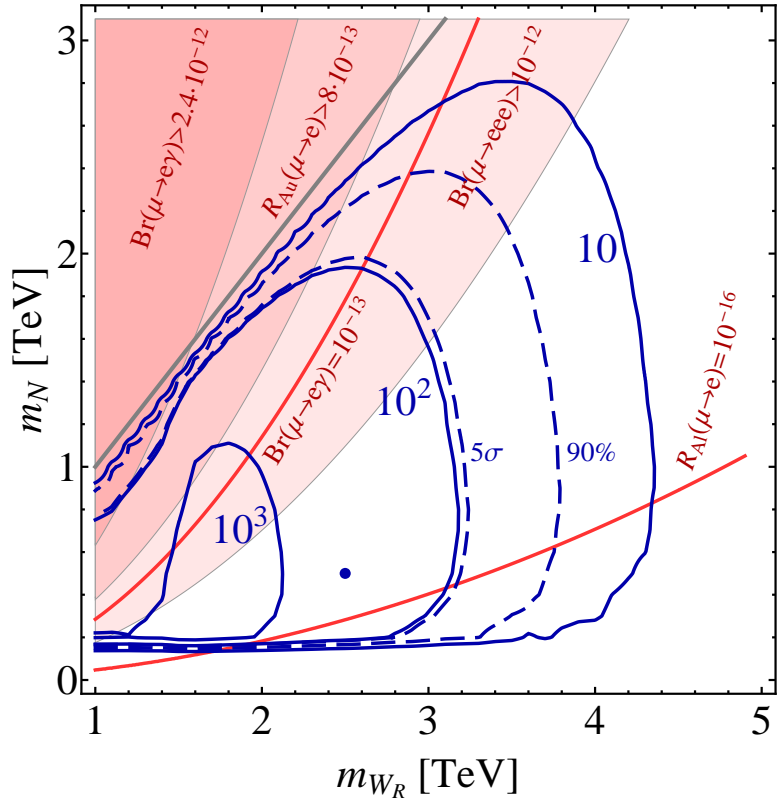}
\caption{Comparison of LFV event rates at the LHC and in low energy rare decays. The solid blue contours give the number of OS+SS events for
  the LFV signature $e^\pm\mu^{\pm,\mp} + 2j$ at the LHC with 14~TeV
  and $\mathcal{L}=30\text{ fb}^{-1}$. The dashed contours
  define the parameter region with signals at $5\sigma$ and
  90\%. The shaded red areas denote the parameter
  regions excluded by current low energy LFV limits, whereas the red
  contours show the expected sensitivity of planned experiments. The
  processes have been calculated using maximal unitary mixing between
  two heavy neutrinos coupling only to $e$ and $\mu$ with a 1\% mass
  splitting, i.e. $\phi=\pi/4$ (cf. Eq.~\ref{eq:twoflavourmix}), $(m_{N_2}-m_{N_1})/m_{N_1} =
  0.01$. The spectrum of the doubly charged Higgs bosons is given by
  $m_{\Delta_{L,R}^{--}} = m_{W_R}$. The grey contour corresponds to the kinematical
  threshold $m_{W_R} = m_N$.}
\label{fig:ComparisonLowEnergy_MWR_MN} 
\end{figure}
Due to the small neutrino three-body decay width, $\Gamma_{N} \propto
g_R^2 m_N^5/m_W^4<10^{-2}$~GeV (cf. Figure~\ref{fig:MW_MN_ee}), the regimes 1) and 2) apply for a wide
range of neutrino mass splittings. In these regimes, the oscillations
between right-handed neutrinos decohere. Consequently, the LFV process
rates at the LHC are not suppressed and are independent of the mass
splitting (unless the individual resonances can be resolved
kinematically), and can be probed at the LHC for mass differences as small as $\Delta m_N \approx \Gamma_N \approx
10^{-2}-10^{-6}$~GeV. This is in stark contrast to low energy rare LFV
processes which, as described in Section~\ref{sec:limitsLFV},
experience a right-handed GIM-like suppression as $\Delta
m^2_N/m^2_{W_R}$. For example, for two flavour $e-\mu$ mixing, the
branching ratio of $\mu\to e\gamma$,
Eq.~(\ref{eq:BrmuegammaSimplified}), can be written in our parametrization as
\begin{equation}
	Br(\mu\to e\gamma) 
	\approx 2 \times 10^{-9} \sin^2(2\phi)
	\left(\frac{\Delta m^2_{12}}{m^2_{W_R}}\right)^2
	\left(\frac{2\text{ TeV}}{m_{W_R}}\right)^4,
\end{equation}
with similar results for the other rare processes. For the
unsuppressed case with large mixing $\phi \approx \pi/4$ and large
mass difference $\Delta m^2_{12} \gtrsim m^2_{W_R}$, the current
limits on this decay already put severe constraints on the scale
$m_{W_R}$. As discussed in Section~\ref{sec:limitsLFV}, the $\mu\to e$
conversion in nuclei and $\mu\to eee$ are expected to restrict the
parameter space even more. On the other hand, even for modest neutrino
mass splittings, e.g. $m_{N_2} - m_{N_1} \approx 50$~GeV in our
benchmark scenario with $m_{N_1} = 500$~GeV and $m_{W_R} = 2000$~GeV,
the branching ratio is already suppressed by a factor $\approx
10^{-4}$ to $Br(\mu\to e\gamma) \approx 4\cdot 10^{-13}$, below the
current experimental limit. The LFV event rate at the LHC would not be
affected by a mass splitting of this size, and it could even be
possible to reconstruct the individual neutrino mass
resonances. Measurements of LFV processes at the LHC and low energy
rare LFV decays therefore provide highly complementary information on
the mass spectrum and the flavour couplings of the LRSM.

\begin{figure}[t!]
\centering
\includegraphics[clip,width=0.49\textwidth]{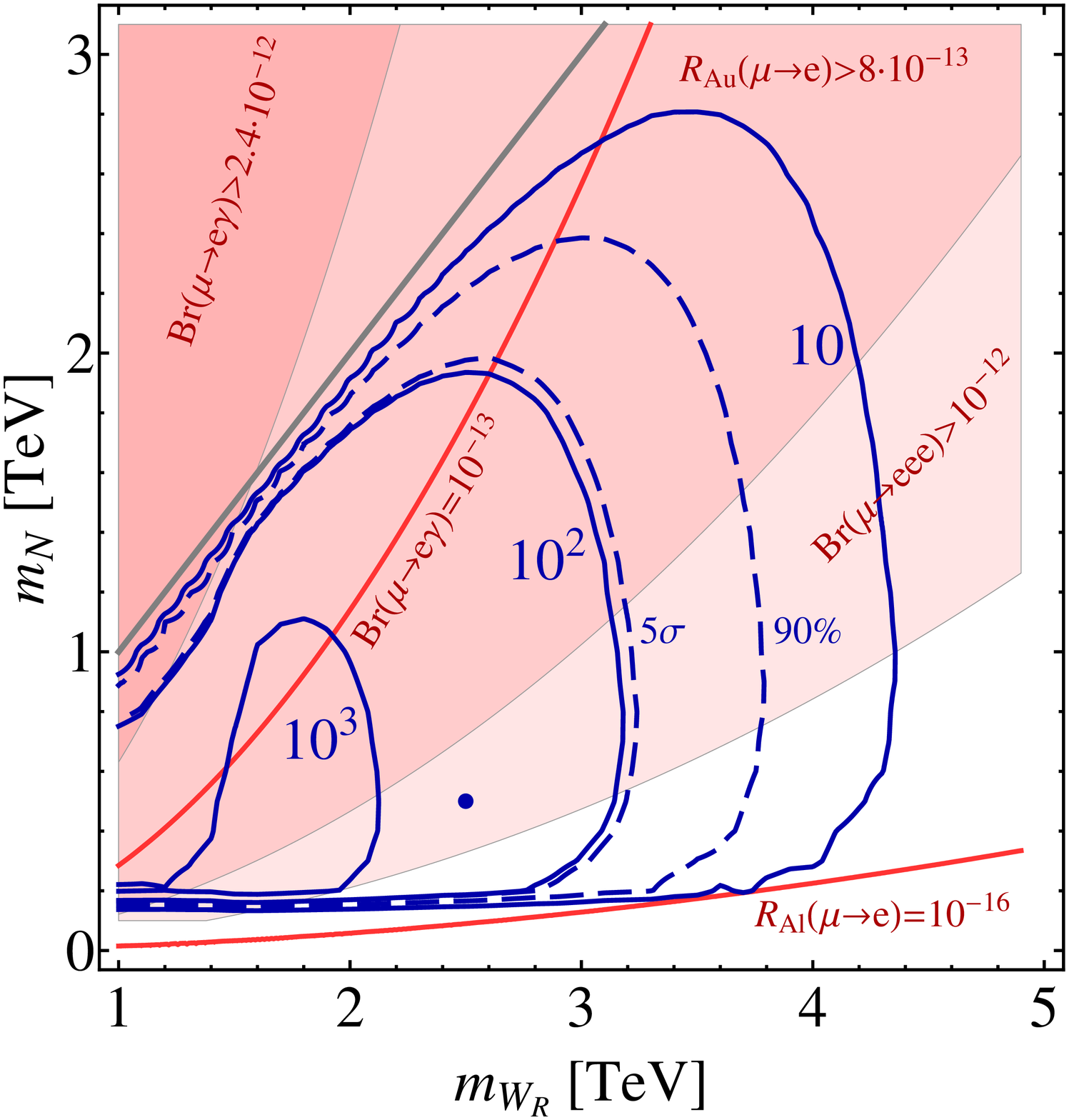}
\includegraphics[clip,width=0.49\textwidth]{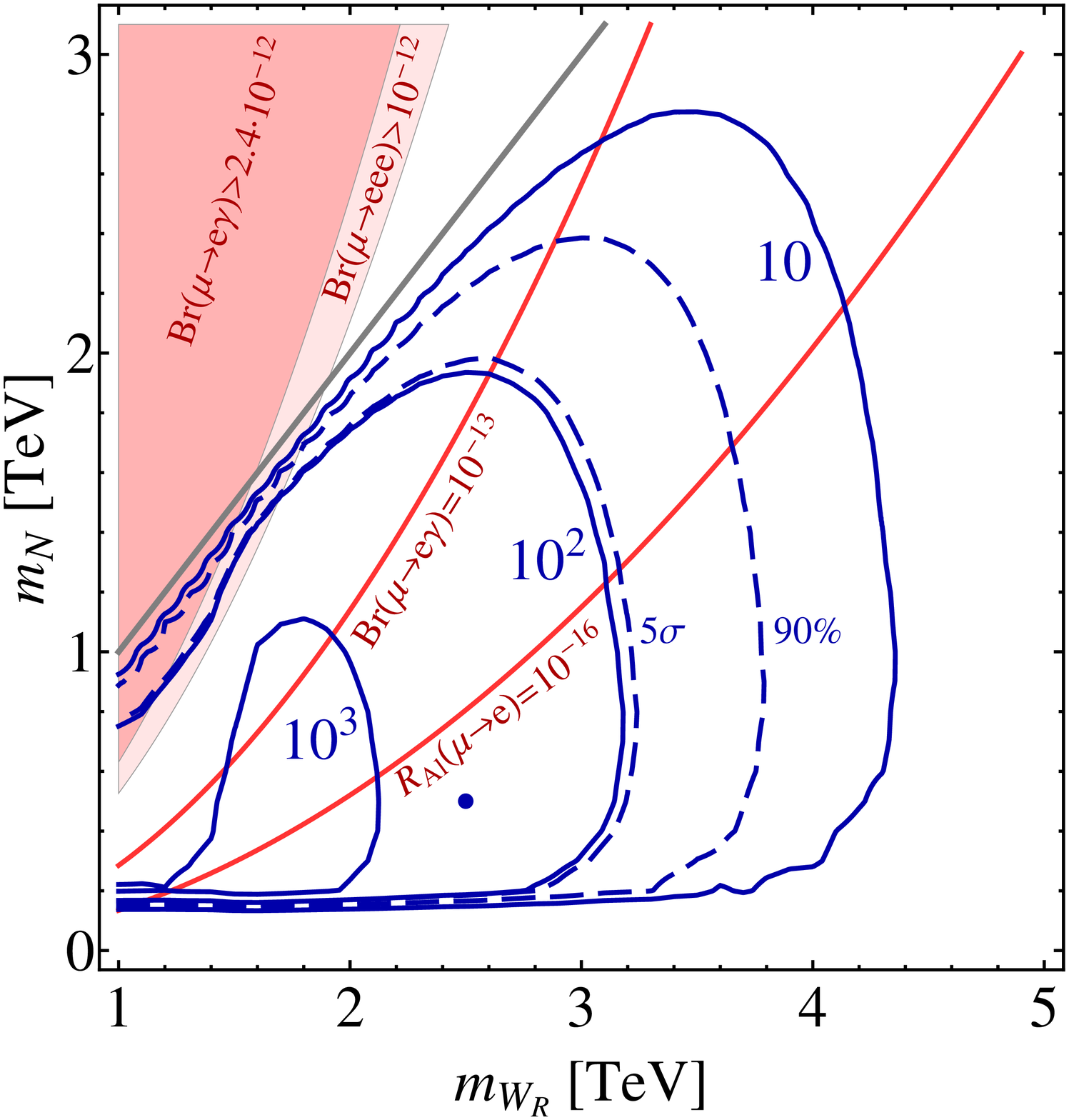}
\caption{As Figure~\ref{fig:ComparisonLowEnergy_MWR_MN}, but with a doubly charged Higgs boson mass spectrum $m_{\Delta_{L,R}^{--}} = 0.3 \times m_{W_R}$ (left) and $m_{\Delta_{L,R}^{--}} = 3\times m_{W_R}$ (right).}
\label{fig:ComparisonLowEnergy_MWR_MN2} 
\end{figure}
To explore this complementarity, we first compare the sensitivity of
$\mu-e$ LFV processes on the masses $m_{W_R}$ and $m_N$ for maximal
mixing $\phi = \pi/4$ and a fixed 1\% neutrino mass splitting,
$(m_{N_2} - m_{N_1})/m_N = 10^{-2}$. This is shown in
Figures~\ref{fig:ComparisonLowEnergy_MWR_MN} and \ref{fig:ComparisonLowEnergy_MWR_MN2} for different heavy Higgs
boson mass spectra: $m_{\Delta_{L,R}^{--}}/m_{W_R} = 0.3, 1, 3$, mapping out the naturally expected
spectral range of scales in the LRSM. As discussed in
Section~\ref{sec:limitsLFV}, the lower the Higgs boson masses are,
the more constraining the processes $\mu-e$ conversion and $\mu\to
eee$ become. The current limits on the rare processes already strongly
constrain the parameter space, with $\mu\to eee$ providing the most
stringent bound. 
Even for $m_{\Delta_{L,R}^{--}}/m_{W_R} = 3$, the
parameter space with $m_N > m_{W_R}$, which is inaccessible at the
LHC, is almost ruled out. For $m_{\Delta_{L,R}^{--}}/m_{W_R} = 0.3$ the
parameter space that could be probed at the LHC is almost ruled out,
and only the region with rather low neutrino masses, $m_N \approx
200-600$~GeV, is still allowed. In this mass regime, the expected
COMET sensitivity on $R^{Al}(\mu\to e) = 10^{-16}$ will fully probe
the LHC accessible parameter space. Generically, there is a high
potential that low energy LFV processes can probe the parameter space
that can be also tested at the LHC, allowing for a highly detailed
view of the LRSM mass spectrum and flavour mixing properties.

\begin{figure}[t!]
\centering
\includegraphics[clip,width=0.62\textwidth]{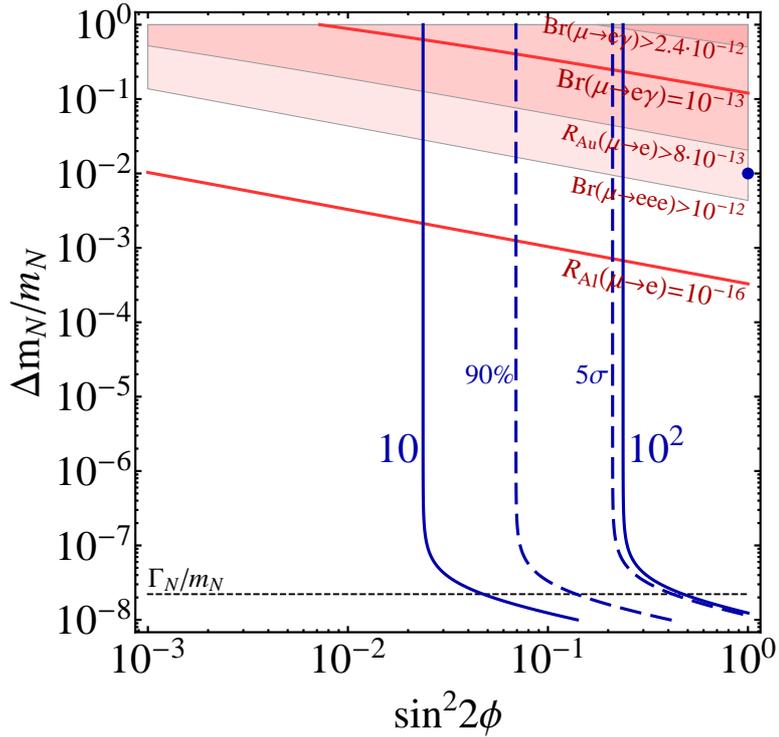}
\caption{As Figure~\ref{fig:ComparisonLowEnergy_MWR_MN}, but showing
  the dependence on the heavy neutrino mixing angle parameter $\sin^2 2\phi$ and the mass splitting $(m_{N_2}-m_{N_1})/m_{N_1}$. The mass
  scales are given by $m_{W_R} = 2.5$~TeV and $m_{N_1} = 0.5$~TeV, with
  a light heavy Higgs sector, $m_{\Delta_{L,R}^{--}} =
  0.3 \times m_{W_R}$. The dashed horizontal line denotes the value
  for the heavy neutrino width $\Gamma_N / m_N$.}
\label{fig:ComparisonLowEnergy_phi_dm} 
\end{figure}
To explore the complementarity in probing different
neutrino mass splittings and flavour mixing, we compare the
sensitivity of low energy and LHC processes on $\sin^2(2\phi)$ and
$\Delta m_N / m_N$ in
Figure~\ref{fig:ComparisonLowEnergy_phi_dm}. Here, the LRSM mass
spectrum is fixed to $m_N = 0.5$~TeV, $m_{W_R} = 2.5$~TeV and
$m_{\Delta_{L,R}^{--}} = 0.66$~TeV, corresponding to a light Higgs
sector. As discussed above, the LFV process rate at the LHC is
independent of the neutrino mass splitting until it becomes comparable
and smaller than the heavy neutrino decay width at $\Gamma_N / m_N
\approx 5\cdot 10^{-8}$. It is therefore possible to probe such tiny
mass differences at the LHC for mixing angles $\sin^2(2\phi) \gsim 10^{-1}$ in this scenario. On the other hand, the low energy
processes exhibit the typical dependence $\propto \sin^2(2\phi)(\Delta
m_N^2)^2$, and may only probe mass splittings as low as $\Delta m_N /
m_N \approx 10^{-3}$.

\subsection{Lepton Number Violation}
\label{sec:LeptonNumberViolation}

So far we have only considered OS and OS+SS event signatures, but not the SS sample independently, as the significance of such a signal depends crucially on a proper treatment of the same-sign lepton background arising from charge mis-identification, mistakenly reconstructed hard leptons from jets and diboson production. Such a treatment goes beyond the scope of this work. Same-sign lepton events are of course a crucial consequence of the Majorana nature of heavy neutrinos in left-right symmetric models, and the associated lepton number violation. In the LRSM considered here, the heavy neutrinos acquire their Majorana masses through the breaking of both the left-right and lepton number symmetry at the high scale $v_R$, much larger than the neutrino Dirac mass terms, see Eq.~(\ref{eq:sub}).

As a result, processes mediated by the heavy neutrinos exhibit maximal lepton number violation, with equal probabilities for the heavy neutrinos decaying into positively and negatively charged leptons. Such a scenario is technically natural to describe both light neutrino masses $\approx 0.1$~eV and heavy neutrino masses $\approx 1$~TeV, if one assumes \emph{ad hoc} small Dirac couplings between light and heavy neutrinos.
On the other hand, accounting for the light neutrino masses within the seesaw mechanism with heavy neutrinos close to the electroweak scale would require the breaking of lepton number at a much lower scale $\mu_\text{LNV}$, as in the inverse or linear seesaw schemes~\cite{mohapatra:1986bd, akhmedov:1995vm, akhmedov:1995ip, malinsky:2005bi, Bazzocchi:2009kc}, or by invoking a proper flavour symmetry among the lepton Yukawa couplings, see e.g. \cite{Gluza:2002vs, pilaftsis:2003gt, Kersten:2007vk, Deppisch:2010fr}.
These scenarios however lead naturally to quasi-Dirac heavy neutrinos where all lepton number violating processes are suppressed by the small mass splitting of heavy neutrinos, $\mu_\text{LNV}/m_N$. From the model building viewpoint it is therefore not straightforward to predict the rate of same-sign lepton events for a given heavy neutrino mass, whereas the rate of events with opposite-sign but different flavours are generally independent of the mechanism of lepton number violation. This question will be addressed in an upcoming analysis.

Despite the above caveats, for illustration we compare the sensitivity of observing lepton number violation at the LHC with that of $0\nu\beta\beta$ experiments in the LRSM. As pointed out in \cite{Cirigliano:2004tc, Tello:2010am}, when staying within the LRSM, it is possible to explore the interplay between LFV and LNV further. The LHC analysis in this section is the same as described in Section~\ref{sec:simulation}, using identical cuts and reconstruction criteria. As illustrated in Table~\ref{tab:CutFlow}, we estimate the same-sign dilepton background by assuming a 5\% probability of mis-identifying the charge of a lepton of an OS event.

\begin{figure}[t!]
\centering
\includegraphics[clip,width=0.62\textwidth]{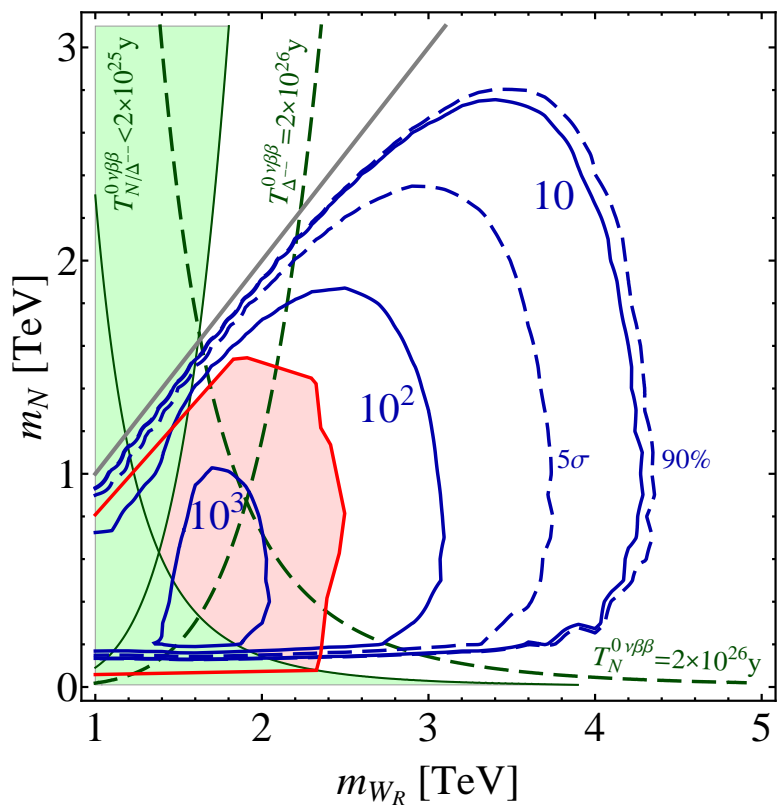}
\caption{Comparison of LNV event rates at the LHC and in $0\nu\beta\beta$ experiments. The solid blue contours give the number of same-sign events for the LNV signature $e^\pm e^\pm + 2j$ at the LHC with 14~TeV and $\mathcal{L}=30\text{ fb}^{-1}$. The dashed blue contours define the parameter regions with a significance at $5\sigma$ and 90\%, respectively, where the background is estimated using a 5\% charge mis-identification probability (see discussion of Table~\ref{tab:CutFlow}). The shaded green area denotes the parameter space excluded by $0\nu\beta\beta$ at $T^{0\nu\beta\beta}\approx 2\times 10^{25}$~yrs, assuming dominant doubly-charged Higgs or heavy neutrino exchange. Correspondingly, the green dashed contours show the sensitivity of future $0\nu\beta\beta$ experiments at $T^{0\nu\beta\beta}\approx 2\times 10^{26}$~yrs. The red shaded area is excluded by current LHC searches at ATLAS~\cite{CERN-PH-EP-2012-022}.}
\label{fig:ComparisonDBD_MWR_MN} 
\end{figure}
In Figure~\ref{fig:ComparisonDBD_MWR_MN} we show the event rate of the same-sign signature $e^\pm e^\pm + 2j$ at the LHC with 14~TeV and
$\mathcal{L}=30\text{ fb}^{-1}$. For comparison, the shaded
green regions and green dashed contours represent the current limit
from $T^{0\nu\beta\beta} \gtrsim 2\times 10^{25}$~yrs
(Heidelberg-Moscow) and the future sensitivity $T^{0\nu\beta\beta} \approx
2\times 10^{26}$~yrs (improvement by one order of magnitude) of
$0\nu\beta\beta$ experiments. Here, we assume that
neutrinoless double beta decay is either dominated by heavy neutrino
exchange (Figure~\ref{fig:diagrams_0nubb}(b) and described by the
effective coupling in Eq.~(\ref{eq:epsilonN})) or by Higgs triplet
exchange (Figure~\ref{fig:diagrams_0nubb}(e) and described by the
effective coupling in Eq.~(\ref{eq:epsilonDelta})). In the latter case,
a light doubly charged Higgs boson mass spectrum is assumed, with
$m_{\Delta^{--}_{L,R}} = 0.33\times m_{W_R}$. In both cases any other
contribution is assumed to be zero. As discussed in
Section~\ref{sec:limits0nubb}, this is naturally the case for the
left-right mixing contributions Figure~\ref{fig:diagrams_0nubb}(c) and
(d), though the contribution from light neutrino exchange
(Figure~\ref{fig:diagrams_0nubb}(a)) will always be present. Hence the
sensitivities shown here would correspond to an hierarchical
light neutrino mass spectrum, with a small effective mass
$m_{\beta\beta}$. Due to the uncertainties described above and because
of the additional dependence on the doubly charged Higgs mass and the
heavy neutrino mixing matrix (here we neglect lepton flavour
violation, i.e. we use $V_{Ne}=1, V_{N\mu} = V_{N\tau} = 0$), the sensitivities shown here are for illustration only. Nevertheless, it is interesting that the area currently probed by the LHC will also be tested in upcoming neutrinoless double beta decay experiments.

\section{Conclusions}
\label{sec:conclusion}

Light neutrino masses naturally arise in left-right symmetric seesaw extensions of the Standard Model, as required in order to account for current neutrino oscillations data.
Lepton flavour violating effects involving charged leptons are also naturally  expected in such scenarios if the masses of the heavy right-handed neutrinos present in left-right symmetric models are of the order of 1 - 10~TeV. Here we have considered lepton flavour violating processes induced in the production and decay of heavy right-handed neutrinos and the resulting signatures at the LHC.
Either through the assumption of small Yukawa couplings or suitable model constructions implementing a low-scale seesaw, for example inverse or linear seesaw, right-handed neutrinos can have masses of order TeV, and are hence accessible at the LHC.  
For this case we have derived the expected LHC sensitivities on the right-handed gauge boson and heavy neutrino masses, as well as the LFV couplings of the heavy neutrinos to charged leptons and compared the collider results with existing bounds from low energy LFV rare decays.
Our discussion was mainly devoted to the first two leptonic flavours, due to their cleaner detection prospects, though extension to the tau flavour will be important, in view of good tau detection efficiencies in decays of heavy particles at the LHC~\cite{AguilarSaavedra:2012fu}.

Because right-handed neutrinos can be produced at gauge coupling strength in left-right symmetric models, the LHC has the potential to discover right-handed $W_R$ bosons up to $m_{W_R} \approx 3-4$~TeV and heavy neutrinos  up to $m_{N_R} \approx 2-2.5$~TeV, at 14~TeV and $\mathcal{L} = 30\text{ fb}^{-1}$. No signal has been found so far, and current bounds from LHC searches are already stringent, with the excluded area extending to $(m_{W_R}, m_{N_R}) \approx (2.5, 1.5)$~TeV. Outside these limits, LFV couplings, described by the heavy neutrino mixing matrix $V$ entering the right-handed charged current interaction $\frac{g_R}{\sqrt{2}} V_{N\ell} \bar N_R \gamma^\mu \ell_R$, as low as $V_{N\ell} \approx 0.3$ can still be probed at the LHC. We have also explored the complementarity of such searches with LFV probes at low energies, namely $\mu\to e\gamma$, $\mu\to 3e$ and $\mu\to e$ conversion in nuclei. If the mass splitting of the heavy right-handed neutrinos is large, $\Delta m_N^2 / m^2_{W_R} \gtrsim 1$, these processes already heavily constrain the presence of LFV in the right-handed neutrino sector. On the other hand, if the mass splittings are small, $\Delta m_N^2 / m^2_{W_R} \lesssim 0.01$, the low energy LFV processes are GIM suppressed whereas LFV can still be observed at the LHC through the resonant production of right-handed neutrinos.
This analysis provides an example of the general complementarity between LFV searches at the LHC and at low energies. Being based on high intensity experiments, low energy probes such as $\mu \to e \gamma$ have a further reach to higher scales of new physics as well as smaller LFV couplings and mixing angles. On the other hand, collider searches are limited by the available energy and luminosity, but have the potential to probe individual particles and couplings.

\begin{figure}[t]
\centering
\includegraphics[clip,width=0.60\textwidth]{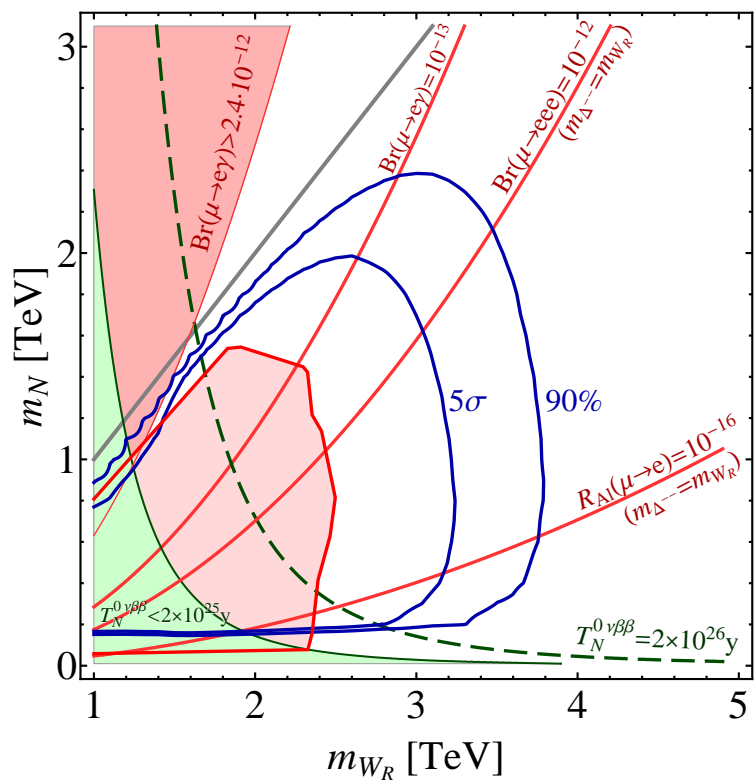}
\caption{Comparison of LFV and LNV event rates at the LHC and in low energy probes. The solid blue contours define the parameter region with signals of the LFV OS+SS process $pp \to W_R \to N_R \ell \to e \mu + 2\text{ jets}$ at the LHC at $5\sigma$ and 90\% (14~TeV, $\mathcal{L} = 30\text{ fb}^{-1}$). The solid gray contour corresponds to the kinematical threshold $m_{W_R}=m_N$. Overlayed are the current and expected future sensitivities of low energy LFV processes and $0\nu\beta\beta$ (mediated by heavy neutrinos), as denoted in the plot. All processes were calculated assuming maximal flavour mixing of two heavy neutrinos $N_1$ and $N_2$ to electrons and muons with a mass difference $m_{N_2}-m_{N_1}=0.01m_N$. The red shaded area in the lower left corner is excluded by current LHC searches at ATLAS.}
\label{fig:MW_MN_overview} 
\end{figure}
In this work, our focus has been on lepton flavour violating effects and the potential of the LHC to probe the flavour mixing of the heavy neutrinos in left-right symmetric models. The dedicated analysis of lepton number violating effects at the LHC requires a thorough simulation of the relevant same-sign dilepton background. On the theoretical side, it also requires a detailed specification of the lepton number symmetry breaking mechanism. Within the minimal left-right symmetric model, lepton number is broken at a high scale, generating the heavy Majorana masses of the right-handed neutrinos. The resulting LNV effects at the LHC are therefore maximal, with the heavy neutrinos decaying with equal probabilities into positive and negative leptons. An overview of the sensitivities of high and low energy probes of LFV and LNV processes in this scenario is shown in Figure~\ref{fig:MW_MN_overview}.

Especially with respect to the origin of lepton number violation, the scenario analyzed in this work is not unique, and there are theoretical consideration to address some of the issues of the minimal left-right symmetric model. Right-handed neutrinos are the messengers whose exchange yields neutrino masses through the type-I seesaw mechanism. Similarly, heavy scalar triplet exchange induces neutrino masses through the type-II seesaw.
It is therefore expected that, at some level, the smallness of neutrino masses will make it difficult, if not preclude, to probe the physics of the heavy right-handed neutrinos.
First note that having TeV-scale right-handed gauge bosons in the minimal \lr model discussed in Sec.~\ref{sec:minimalLR} does not, by itself, provide a fully satisfactory picture.
For example, the gauge couplings in such minimal scheme have no simple SO(10) embedding compatible with gauge coupling unification. On the other hand right-handed neutrinos at the TeV scale can only be possible through the \textit{ad-hoc} requirement of tiny Dirac neutrino Yukawa couplings in Eq.~(\ref{eq:Lagrangian}).

These shortcomings can all be naturally evaded by implementing a low scale seesaw mechanism, such as inverse~\cite{mohapatra:1986bd, Bazzocchi:2009kc} or linear seesaw~\cite{malinsky:2005bi} within the \lr context~\cite{akhmedov:1995ip, akhmedov:1995vm}. This would not only justify the lightness of the right-handed neutrinos, without the need to invoke unnaturally small Yukawa couplings, but it also achieves a consistent gauge coupling unification within supersymmetry~\cite{DeRomeri:2011ie}.

\section*{Acknowledgements}

We thank Stephen Bieniek, Martin Hirsch and Mikael Rodrigu\'ez for fruitful discussions. We would also like to thank Francisco del Aguila and Juan Antonio Aguilar-Saavedra for a careful reading of the manuscript and useful comments. This work was supported by the Spanish MEC under grants FPA2008-00319/FPA,
FPA2011-22975 and MULTIDARK CSD2009-00064 (Consolider\-Ingenio 2010 Programme), by Prometeo/2009/091 (Generalitat Valenciana), by the EU ITN UNILHC PITN-GA-2009-237920. The work of O.K. has been supported by a CPAN fellowship. S.P.D. also acknowledged financial support from DST, Government of India, SR/MF/PS-03/2009-VB-I.

\bibliographystyle{h-physrev4}
\bibliography{merged}

\end{document}